\documentclass[%
 reprint,
superscriptaddress,
showpacs,
amsmath,amssymb,
aps,
]{revtex4-1}

\usepackage{graphicx}% Include figure files
\usepackage{dcolumn}% Align table columns on decimal point
\usepackage{bm}% bold math
\usepackage{amsmath}
\usepackage{autobreak}
\usepackage{hyperref}% add hypertext capabilities
\usepackage[mathlines]{lineno}% Enable numbering of text and display math
\usepackage{natbib}
\usepackage{multirow}
\usepackage{booktabs}
\usepackage{mathrsfs}

\begin{document}

\preprint{APS/123-QED}

\title{Search for Cosmic Ray Electron Boosted Dark Matter with the CDEX-10 Experiment}

\author{R.~Xu}
\affiliation{Key Laboratory of Particle and Radiation Imaging (Ministry of Education) and Department of Engineering Physics, Tsinghua University, Beijing 100084}
\author{L.~T.~Yang}\altaffiliation [Corresponding author: ]{yanglt@mail.tsinghua.edu.cn}
\affiliation{Key Laboratory of Particle and Radiation Imaging (Ministry of Education) and Department of Engineering Physics, Tsinghua University, Beijing 100084}
\author{Q. Yue}\altaffiliation [Corresponding author: ]{yueq@mail.tsinghua.edu.cn}
\affiliation{Key Laboratory of Particle and Radiation Imaging (Ministry of Education) and Department of Engineering Physics, Tsinghua University, Beijing 100084}

\author{K.~J.~Kang}
\affiliation{Key Laboratory of Particle and Radiation Imaging (Ministry of Education) and Department of Engineering Physics, Tsinghua University, Beijing 100084}
\author{Y.~J.~Li}
\affiliation{Key Laboratory of Particle and Radiation Imaging (Ministry of Education) and Department of Engineering Physics, Tsinghua University, Beijing 100084}

\author{H.~P.~An}
\affiliation{Key Laboratory of Particle and Radiation Imaging (Ministry of Education) and Department of Engineering Physics, Tsinghua University, Beijing 100084}
\affiliation{Department of Physics, Tsinghua University, Beijing 100084}

\author{Greeshma~C.}
\altaffiliation{Participating as a member of TEXONO Collaboration}
\affiliation{Institute of Physics, Academia Sinica, Taipei 11529}

\author{J.~P.~Chang}
\affiliation{NUCTECH Company, Beijing 100084}
\author{H.~Chen}
\affiliation{Key Laboratory of Particle and Radiation Imaging (Ministry of Education) and Department of Engineering Physics, Tsinghua University, Beijing 100084}

\author{Y.~H.~Chen}
\affiliation{YaLong River Hydropower Development Company, Chengdu 610051}
\author{J.~P.~Cheng}
\affiliation{Key Laboratory of Particle and Radiation Imaging (Ministry of Education) and Department of Engineering Physics, Tsinghua University, Beijing 100084}
\affiliation{School of Physics and Astronomy, Beijing Normal University, Beijing 100875}
\author{J.~Y.~Cui}
\affiliation{Key Laboratory of Particle and Radiation Imaging (Ministry of Education) and Department of Engineering Physics, Tsinghua University, Beijing 100084}
\author{W.~H.~Dai}
\affiliation{Key Laboratory of Particle and Radiation Imaging (Ministry of Education) and Department of Engineering Physics, Tsinghua University, Beijing 100084}
\author{Z.~Deng}
\affiliation{Key Laboratory of Particle and Radiation Imaging (Ministry of Education) and Department of Engineering Physics, Tsinghua University, Beijing 100084}
\author{Y.~X.~Dong}
\affiliation{Key Laboratory of Particle and Radiation Imaging (Ministry of Education) and Department of Engineering Physics, Tsinghua University, Beijing 100084}
\author{C.~H.~Fang}
\affiliation{College of Physics, Sichuan University, Chengdu 610065}

\author{H.~Gong}
\affiliation{Key Laboratory of Particle and Radiation Imaging (Ministry of Education) and Department of Engineering Physics, Tsinghua University, Beijing 100084}
\author{Q.~J.~Guo}
\affiliation{School of Physics, Peking University, Beijing 100871}
\author{T.~Guo}
\affiliation{Key Laboratory of Particle and Radiation Imaging (Ministry of Education) and Department of Engineering Physics, Tsinghua University, Beijing 100084}
\author{X.~Y.~Guo}
\affiliation{YaLong River Hydropower Development Company, Chengdu 610051}
\author{L.~He}
\affiliation{NUCTECH Company, Beijing 100084}
\author{J.~R.~He}
\affiliation{YaLong River Hydropower Development Company, Chengdu 610051}

\author{H.~X.~Huang}
\affiliation{Department of Nuclear Physics, China Institute of Atomic Energy, Beijing 102413}
\author{T.~C.~Huang}
\affiliation{Sino-French Institute of Nuclear and Technology, Sun Yat-sen University, Zhuhai 519082}

\author{S.~Karmakar}
\altaffiliation{Participating as a member of TEXONO Collaboration}
\affiliation{Institute of Physics, Academia Sinica, Taipei 11529}

\author{Y.~S.~Lan}
\affiliation{Key Laboratory of Particle and Radiation Imaging (Ministry of Education) and Department of Engineering Physics, Tsinghua University, Beijing 100084}
\author{H.~B.~Li}
\altaffiliation{Participating as a member of TEXONO Collaboration}
\affiliation{Institute of Physics, Academia Sinica, Taipei 11529}
\author{H.~Y.~Li}
\affiliation{College of Physics, Sichuan University, Chengdu 610065}
\author{J.~M.~Li}
\affiliation{Key Laboratory of Particle and Radiation Imaging (Ministry of Education) and Department of Engineering Physics, Tsinghua University, Beijing 100084}
\author{J.~Li}
\affiliation{Key Laboratory of Particle and Radiation Imaging (Ministry of Education) and Department of Engineering Physics, Tsinghua University, Beijing 100084}
\author{M.~C.~Li}
\affiliation{YaLong River Hydropower Development Company, Chengdu 610051}
\author{Q.~Y.~Li}
\affiliation{College of Physics, Sichuan University, Chengdu 610065}
\author{R.~M.~J.~Li}
\affiliation{College of Physics, Sichuan University, Chengdu 610065}
\author{X.~Q.~Li}
\affiliation{School of Physics, Nankai University, Tianjin 300071}
\author{Y.~L.~Li}
\affiliation{Key Laboratory of Particle and Radiation Imaging (Ministry of Education) and Department of Engineering Physics, Tsinghua University, Beijing 100084}
\author{Y.~F.~Liang}
\affiliation{Key Laboratory of Particle and Radiation Imaging (Ministry of Education) and Department of Engineering Physics, Tsinghua University, Beijing 100084}
\author{B.~Liao}
\affiliation{School of Physics and Astronomy, Beijing Normal University, Beijing 100875}
\author{F.~K.~Lin}
\altaffiliation{Participating as a member of TEXONO Collaboration}
\affiliation{Institute of Physics, Academia Sinica, Taipei 11529}
\author{S.~T.~Lin}
\affiliation{College of Physics, Sichuan University, Chengdu 610065}
\author{J.~X.~Liu}
\affiliation{Key Laboratory of Particle and Radiation Imaging (Ministry of Education) and Department of Engineering Physics, Tsinghua University, Beijing 100084}
\author{R.~Z.~Liu}
\affiliation{Key Laboratory of Particle and Radiation Imaging (Ministry of Education) and Department of Engineering Physics, Tsinghua University, Beijing 100084}
\author{S.~K.~Liu}
\affiliation{College of Physics, Sichuan University, Chengdu 610065}
\author{Y.~D.~Liu}
\affiliation{School of Physics and Astronomy, Beijing Normal University, Beijing 100875}
\author{Y.~Liu}
\affiliation{College of Physics, Sichuan University, Chengdu 610065}
\author{Y.~Y.~Liu}
\affiliation{School of Physics and Astronomy, Beijing Normal University, Beijing 100875}
\author{H.~Ma}
\affiliation{Key Laboratory of Particle and Radiation Imaging (Ministry of Education) and Department of Engineering Physics, Tsinghua University, Beijing 100084}
\author{Y.~C.~Mao}
\affiliation{School of Physics, Peking University, Beijing 100871}
\author{A.~Mureed}
\affiliation{College of Physics, Sichuan University, Chengdu 610065}
\author{H.~Pan}
\affiliation{NUCTECH Company, Beijing 100084}
\author{N.~C.~Qi}
\affiliation{YaLong River Hydropower Development Company, Chengdu 610051}
\author{J.~Ren}
\affiliation{Department of Nuclear Physics, China Institute of Atomic Energy, Beijing 102413}
\author{X.~C.~Ruan}
\affiliation{Department of Nuclear Physics, China Institute of Atomic Energy, Beijing 102413}
\author{M.~B.~Shen}
\affiliation{YaLong River Hydropower Development Company, Chengdu 610051}
\author{H.~Y.~Shi}
\affiliation{College of Physics, Sichuan University, Chengdu 610065}
\author{M.~K.~Singh}
\altaffiliation{Participating as a member of TEXONO Collaboration}
\affiliation{Institute of Physics, Academia Sinica, Taipei 11529}
\affiliation{Department of Physics, Banaras Hindu University, Varanasi 221005}
\author{T.~X.~Sun}
\affiliation{School of Physics and Astronomy, Beijing Normal University, Beijing 100875}
\author{W.~L.~Sun}
\affiliation{YaLong River Hydropower Development Company, Chengdu 610051}
\author{C.~J.~Tang}
\affiliation{College of Physics, Sichuan University, Chengdu 610065}
\author{Y.~Tian}
\affiliation{Key Laboratory of Particle and Radiation Imaging (Ministry of Education) and Department of Engineering Physics, Tsinghua University, Beijing 100084}
\author{H.~F.~Wan}
\affiliation{Key Laboratory of Particle and Radiation Imaging (Ministry of Education) and Department of Engineering Physics, Tsinghua University, Beijing 100084}
\author{G.~F.~Wang}
\affiliation{School of Physics and Astronomy, Beijing Normal University, Beijing 100875}
\author{J.~Z.~Wang}
\affiliation{Key Laboratory of Particle and Radiation Imaging (Ministry of Education) and Department of Engineering Physics, Tsinghua University, Beijing 100084}
\author{L.~Wang}
\affiliation{School of Physics and Astronomy, Beijing Normal University, Beijing 100875}
\author{Q.~Wang}
\affiliation{College of Physics, Sichuan University, Chengdu 610065}
\author{Q.~Wang}
\affiliation{Key Laboratory of Particle and Radiation Imaging (Ministry of Education) and Department of Engineering Physics, Tsinghua University, Beijing 100084}
\affiliation{Department of Physics, Tsinghua University, Beijing 100084}
\author{Y.~F.~Wang}
\affiliation{Key Laboratory of Particle and Radiation Imaging (Ministry of Education) and Department of Engineering Physics, Tsinghua University, Beijing 100084}
\author{Y.~X.~Wang}
\affiliation{School of Physics, Peking University, Beijing 100871}
\author{H.~T.~Wong}
\altaffiliation{Participating as a member of TEXONO Collaboration}
\affiliation{Institute of Physics, Academia Sinica, Taipei 11529}

\author{Y.~C.~Wu}
\affiliation{Key Laboratory of Particle and Radiation Imaging (Ministry of Education) and Department of Engineering Physics, Tsinghua University, Beijing 100084}
\author{H.~Y.~Xing}
\affiliation{College of Physics, Sichuan University, Chengdu 610065}
\author{K.~Z.~Xiong}
\affiliation{YaLong River Hydropower Development Company, Chengdu 610051}

\author{Y.~Xu}
\affiliation{School of Physics, Nankai University, Tianjin 300071}
\author{T.~Xue}
\affiliation{Key Laboratory of Particle and Radiation Imaging (Ministry of Education) and Department of Engineering Physics, Tsinghua University, Beijing 100084}
\author{Y.~L.~Yan}
\affiliation{College of Physics, Sichuan University, Chengdu 610065}
\author{N.~Yi}
\affiliation{Key Laboratory of Particle and Radiation Imaging (Ministry of Education) and Department of Engineering Physics, Tsinghua University, Beijing 100084}
\author{C.~X.~Yu}
\affiliation{School of Physics, Nankai University, Tianjin 300071}
\author{H.~J.~Yu}
\affiliation{NUCTECH Company, Beijing 100084}
\author{X.~Yu}
\affiliation{Key Laboratory of Particle and Radiation Imaging (Ministry of Education) and Department of Engineering Physics, Tsinghua University, Beijing 100084}
\author{M.~Zeng}
\affiliation{Key Laboratory of Particle and Radiation Imaging (Ministry of Education) and Department of Engineering Physics, Tsinghua University, Beijing 100084}
\author{Z.~Zeng}
\affiliation{Key Laboratory of Particle and Radiation Imaging (Ministry of Education) and Department of Engineering Physics, Tsinghua University, Beijing 100084}

\author{F.~S.~Zhang}
\affiliation{School of Physics and Astronomy, Beijing Normal University, Beijing 100875}

\author{P.~Zhang}
\affiliation{Key Laboratory of Particle and Radiation Imaging (Ministry of Education) and Department of Engineering Physics, Tsinghua University, Beijing 100084}

\author{P.~Zhang}
\affiliation{YaLong River Hydropower Development Company, Chengdu 610051}

\author{Z.~Y.~Zhang}
\affiliation{Key Laboratory of Particle and Radiation Imaging (Ministry of Education) and Department of Engineering Physics, Tsinghua University, Beijing 100084}

\author{M.~G.~Zhao}
\affiliation{School of Physics, Nankai University, Tianjin 300071}

\author{J.~F.~Zhou}
\affiliation{YaLong River Hydropower Development Company, Chengdu 610051}
\author{Z.~Y.~Zhou}
\affiliation{Department of Nuclear Physics, China Institute of Atomic Energy, Beijing 102413}
\author{J.~J.~Zhu}
\affiliation{College of Physics, Sichuan University, Chengdu 610065}

\collaboration{CDEX Collaboration}
\noaffiliation

\date{\today}

\begin{abstract}
We present new constraints on the cosmic ray electron boosted light dark matter (CReDM) using the 205.4 kg$\cdot$day data of the CDEX-10 experiment located at the China Jinping Underground Laboratory. The cosmic ray electron spectrum and distribution in the Galaxy are generated by the $\tt GALPROP$ code package. In the calculation process of DM-electron scattering process in the Galaxy, we consider the energy-dependency of the DM-electron scattering cross section. The constraints on CReDM are set for both heavy and light mediator scenarios using the CDEX-10 dataset. The result exceeds previous Standard Halo Model (SHM) limits for DM mass lower than 0.6 MeV in heavy mediator case and corresponds to the best sensitivity among all direct detection experiments from 1 keV to 0.5 MeV in the light mediator scenario.
\end{abstract}
\maketitle

\section{\label{sec1}Introduction}
The existence of dark matter (DM, denoted as $\chi$) is supported by numerous cosmological evidences, while no clear signals have been probed yet~\cite{RPP2024,bertone_particle_2005}. Direct detection (DD) experiments such as XENON~\cite{XENON:2025vwd}, LZ~\cite{LZ:2025iaw}, PandaX~\cite{PandaX:2025rrz}, DarkSide~\cite{DarkSide-50:2023fgf}, CRESST~\cite{cresst}, SuperCDMS~\cite{cdmslite}, CoGeNT~\cite{cogent2013} and CDEX~\cite{cdex0,cdex1,cdex12014,cdex12016,cdex1b2018,cdex102018,cdex10_tech,cdex1b_am,cdex10_eft,CDEX_electron,CDEX_black_hole,CDEX_neutrino,CDEX_PBHvDM,CDEX:SRDM,CDEX:supernova}, 
are based on DM-nucleus ($\chi$-$N$) elastic scattering through spin-independent (SI) and spin-dependent interactions. DD experiments lose sensitivity in the sub-GeV DM mass range, as light DM particles cannot transfer sufficient energy to nuclei in detectors to generate signals above the detector threshold. In the DM-electron ($\chi$-$e$) elastic scattering process, light DM particles can potentially pass most of their energies onto electrons, producing observable signals in detectors. Several DD experiments have been dedicated to $\chi$-$e$ scattering searches, including experiments using solid-state detectors such as SENSEI~\cite{SENSEI:2024yyt}, DAMIC~\cite{DAMIC:2023rcc}, EDELWEISS~\cite{EDELWEISS_electron}, SuperCDMS~\cite{supercdms} and CDEX~\cite{CDEX_electron,CDEX:SRDM}, and experiments using liquid-noble gases, such as XENON~\cite{XENON_electron}, PandaX~\cite{pandax_electron}, and DarkSide~\cite{DarkSide-50:2023fgf}.

Recently it has been proposed that light DM particles can be boosted to relativistic or near relativistic momenta by the elastic scattering with cosmic ray (CR) particles in the Milky Way halo~\cite{CRDM_prl,CRDM_TJ,CRDM_ZYF,CRDM_PRD,CRDM_neutrino,CRDM_prd2,PandaX-CRDM,CDEX_CRDM,aziumth_assymetry,COSINE-100:2023tcq,IceCube:BDM,LZ:2025iaw}. The physics channels of $\chi$-$N$ and $\chi$-$e$ boosting have been studied. The cosmic ray boosted DM scenario has been analyzed under the $\chi$-$N$ scattering scenario using the CDEX-10 data in previous works~\cite{CDEX_CRDM}. In this work, we present novel $\chi$-$e$ elastic scattering constraints under the cosmic ray electron boosted DM (CReDM) scenario using the CDEX-10 experiment data, which is the first CReDM constraint given by germanium-based DD experiments.

CDEX-10 experiment~\cite{cdex0,cdex1,cdex12014,cdex12016,cdex1b2018,cdex102018,cdex10_tech,cdex10_eft,cdex1b_am,CDEX_electron} runs a 10-kg $p$-type point contact germanium (PPCGe)~\cite{soma2016} detector array in the China Jinping Underground Laboratory (CJPL) with a rock overburden of 2400 meters~\cite{cjpl}. The detector array is directly immersed in a liquid nitrogen (L$\rm N_2$) cryostat, with 20 cm thick copper and 1 m thick polyethylene serving as shielding. The configuration of the detector system was described in detail previously~\cite{cdex102018,cdex10_tech}. The total exposure reached 205.4 kg$\cdot$day with an analysis threshold of 160 keVee (electron equivalent energy)~\cite{cdex_dark+pthoton}.

This work is organized as follows: firstly we will discuss the cosmic ray electron spectrum in the Galaxy and derive the CReDM spectrum in space, the impacts of the energy dependent cross section are also shown in this section, where the heavy and light mediator case will be discussed; then we will introduce the calculation process of the expected $\chi$-$e$ scattering spectra; finally the exclusion limits will be presented and discussions on the results will be made. 

\section{\label{sec2}CReDM flux in the space} 
In this section, we introduce the CR electron (CRE) spectrum and spatial distribution data used in this analysis. The boosted DM flux is derived and shown based on the CRE inputs.
\subsection{\label{sec_CR}Cosmic ray electron flux in the Galaxy}
Cosmic rays consist of numerous kinds of particles, including protons, electrons, and heavier nuclei. The propagation of CR charged particles in the Galaxy can be modeled by the diffusion model, in which the diffusion halo is parameterized by a cylinder with a radius $R$ of $\sim$ 20 kpc and a half-height $z$ of 1--10 kpc. The CREs are believed to originate from the supernova remnants (SNRs)~\cite{cr_origin,cr_origin2}. The spatial distribution of the primary CRE source $q(R,z)$ follows the standard SNR distribution:
\begin{equation}
    q(R,z)=(\frac{R}{R_{\odot}})^a \mathrm{exp}(-b\frac{R-R_{\odot}}{R_{\odot}})e^{-|z|/z_s},
\end{equation}
where $R_{\odot}=8.5$ kpc is the distance from the Galactic Center (GC) to the Earth. The parameter values $a=1.25$, $b=3.56$ and $z_s=0.2$ kpc are adopted in accordance with previous study~\cite{aziumth_assymetry}. The spectrum of CRE has been measured by numerous space-based experiments, such as AMS~\cite{ams02}, CREAM~\cite{CREAM}, DAMPE~\cite{DAMPE}, CARLET~\cite{CARLET}, and PAMELA~\cite{pamela}. The observed CRE spectrum near the Earth peaks at around 1 GeV due to the solar modulation effect. Voyager I has detected the local interstellar spectrum (LIS) of CREs away from the solar system down to a few MeV, giving results many magnitudes larger than those measured near the Earth~\cite{voyager}.

In this work, we use the $\tt GALPROP$-$\tt v57$ code to generate the required spatial distribution and spectrum of CREs in the Galaxy~\cite{GALPROP_v57}. The propagation parameters reported by former astronomical studies are used in this analysis and listed in Tab.~\ref{tab:table1}. The best-fit values of CRE injection spectral parameters tuned to reproduce the AMS-02~\cite{ams02} and the Voyager I~\cite{voyager} data are also listed.
\begin{table}[!tbp]
\renewcommand\arraystretch{1.2}
\caption{\label{tab:table1}
The best-fit values of propagation parameters and injection spectral parameters from Ref.~\cite{Boschini_2018}.} 
\begin{ruledtabular}
\begin{tabular}{lcc}
  \multicolumn{1}{c}{Parameters}
         &   \multicolumn{1}{c}{Values} 
          \\
\hline
\multirow{1}{*}{$z_h$ (kpc)} & 4.0 kpc   \\
\multirow{1}{*}{$D_0$ ($10^{28} \mathrm{cm^2 s^{-1}}$)} & 4.3  \\
\multirow{1}{*}{$\delta$} & 0.405  \\
\multirow{1}{*}{$V_a$ (km/s)} & 31  \\
\multirow{1}{*}{$dV_c/dz$ ($\mathrm{km\ s^{-1} kpc^{-1}}$)} & 9.8  \\
\multirow{1}{*}{$\rho_0$ (GV)} & 0.19  \\
\multirow{1}{*}{$\rho_1$ (GV)} & 6  \\
\multirow{1}{*}{$\rho_2$ (GV)} & 95  \\
\multirow{1}{*}{$\gamma_0$} & 2.57  \\
\multirow{1}{*}{$\gamma_1$} & 1.40  \\
\multirow{1}{*}{$\gamma_2$} & 2.80  \\
\multirow{1}{*}{$\gamma_3$} & 2.40  \\
\end{tabular}
\end{ruledtabular}
\end{table}

\begin{figure}[!tbp]
\includegraphics[width=\linewidth]{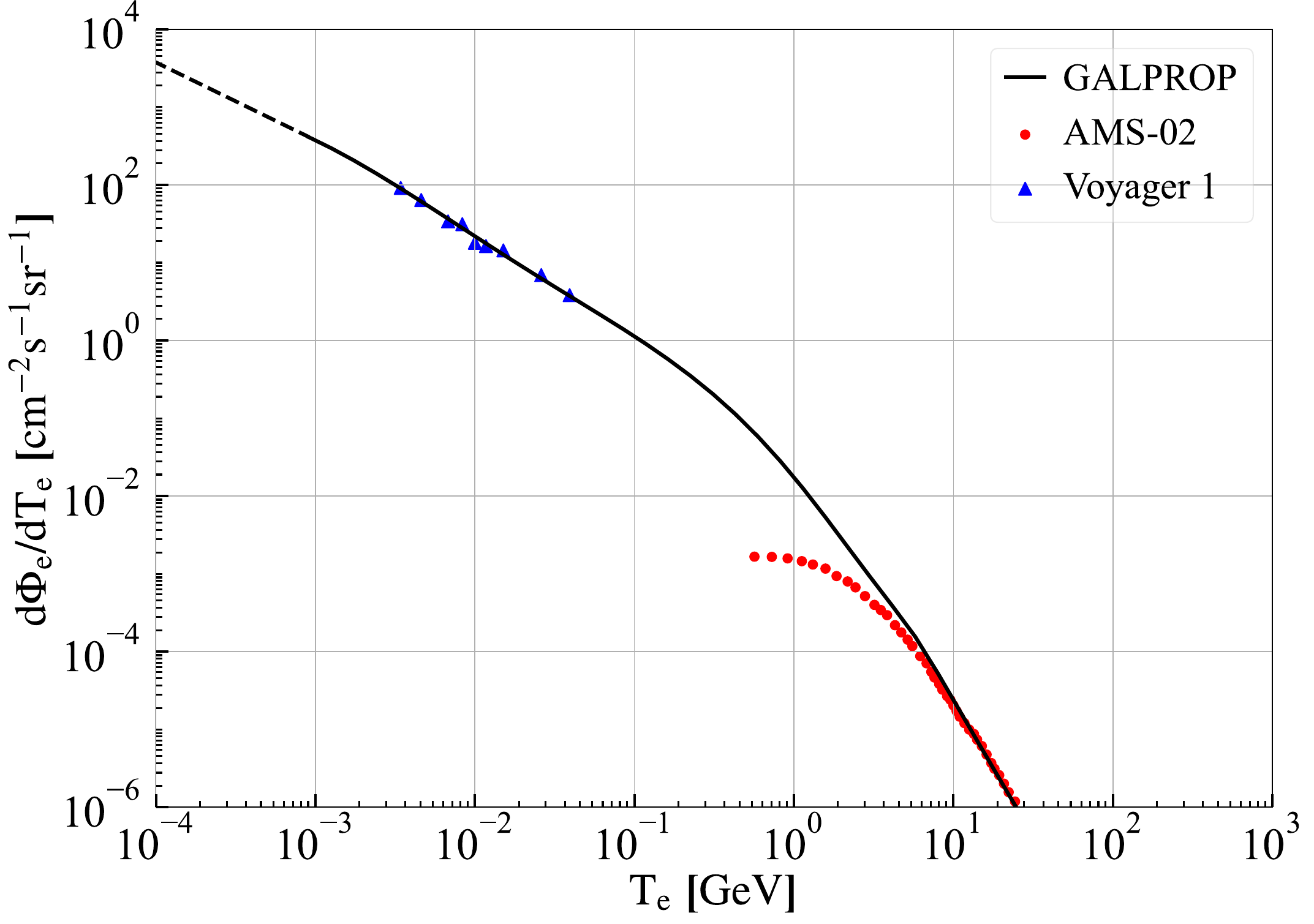}
\caption{The LIS of cosmic ray electrons generated by $\tt GALPROP$ is shown by the black line. The dashed line corresponds to the extrapolation at low energy. The measured data of the AMS-02~\cite{ams02} and Voyager I~\cite{voyager} are shown in red and blue points, respectively.}
\label{fig::cr}
\end{figure}

\begin{figure}[!htbp]
\includegraphics[width=\linewidth]{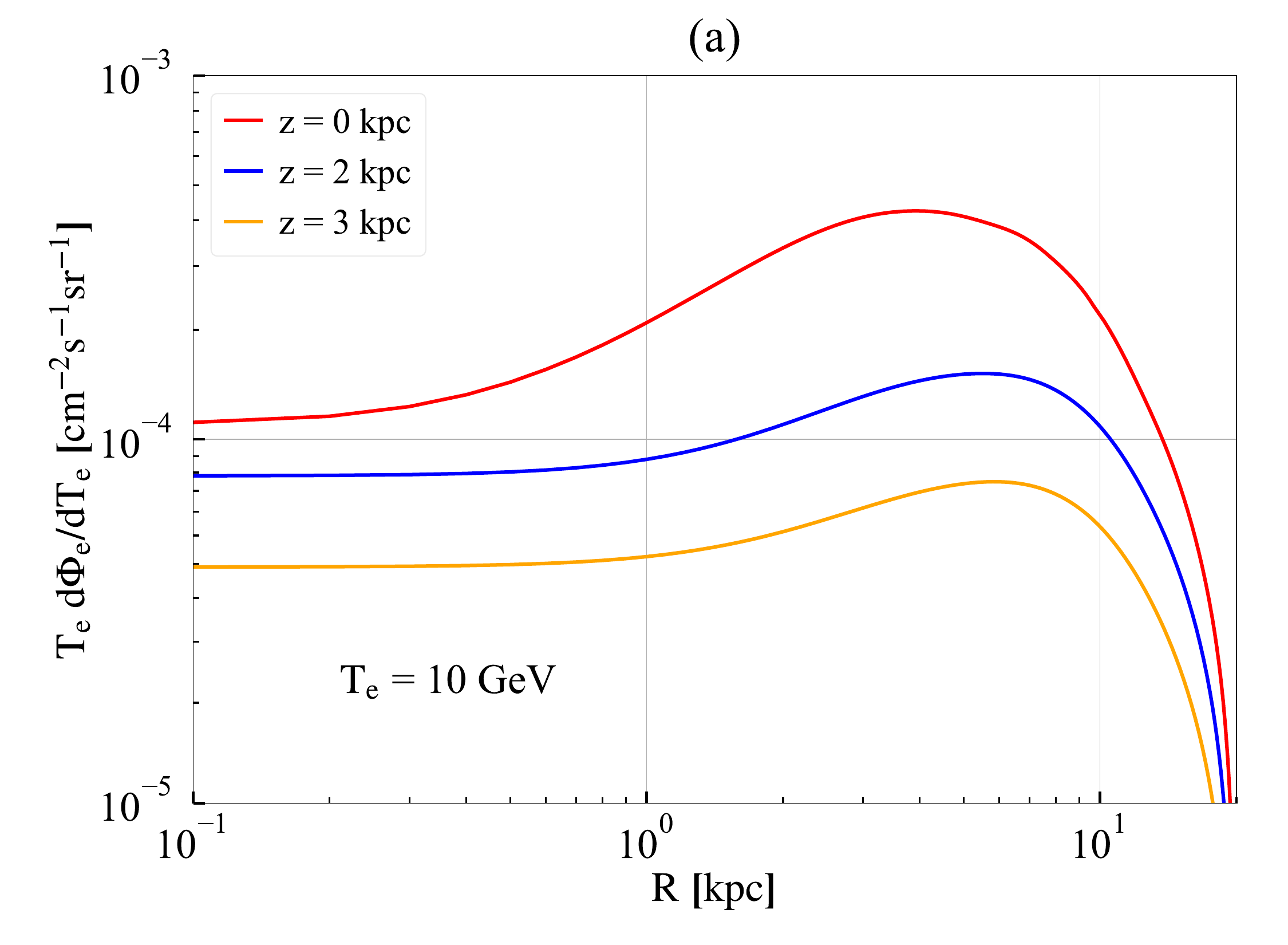}
\includegraphics[width=\linewidth]{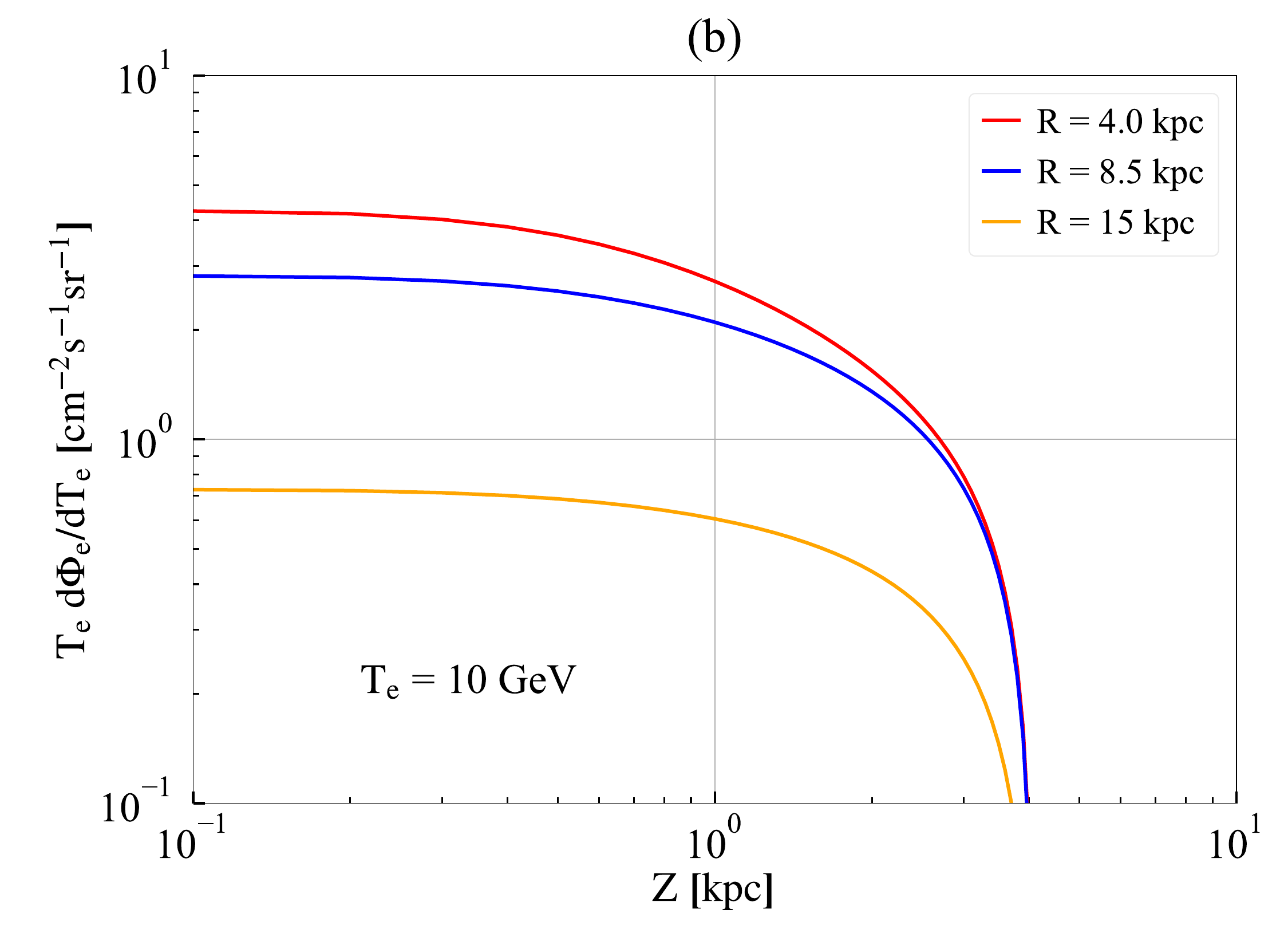}
\caption{(a) The CRE fluxes as a function of distance $R$ from the Galactic Center at different heights $z$ from the Galactic disk. The kinetic energies of the CRE are set to be 10 GeV. (b) The CRE fluxes as a function of height $z$ from the Galactic disk at different radii from the GC. The kinetic energies of the CRE are set to be 10 GeV.}
\label{fig::species}
\end{figure}

The injection spectrum of CRE primary source for $\tt GALPROP$ simulation is assumed to follow a broken power law with three breaks. The breaks $\rho_0-\rho_1$ and exponential indices $\gamma_0-\gamma_3$ are listed in Tab.~\ref{tab:table1}. The low energy part of the CR electron spectrum needs to be inferred indirectly, as there are no measured data of electron LIS below 2 MeV~\cite{voyager}. For the low energy spectrum under 2 MeV, we have adopted a power-law extrapolation of the spectrum as $\propto T_e^{-1}$, which agrees with the studies on the cosmic ray ionization of molecular clouds~\cite{molecular_cloud1,molecular_cloud2}. The energy range of the CRE spectra used in this analysis is from $10^{-10}$ GeV to $10^2$ GeV. The LIS of CRE generated by the $\tt GALPROP$ is shown in Fig.~\ref{fig::cr} together with the measured data of AMS-02 and Voyager I. The spectrum fits well with the measured data of the Voyager I in the low energy region. The spectrum is also in agreement with the AMS-02 measured data in the 10 GeV and higher energy region.

The CRE flux was regarded as homogeneous in the Galaxy in former CReDM analyses~\cite{CRDM_prl,CRDM_PRD}, which is an oversimplified assumption. The CRE flux spatial distribution in a cylinder of radius 20 kpc and half-height 4 kpc is produced using the $\tt GALPROP$ code. In Fig.~\ref{fig::species}, we show the dependence of the CRE flux on the radius $R$ from the GC and the height $z$ at the kinetic energy 10 GeV. The CRE flux increases slowly with $R$ and peaks at about 4 kpc, and then drops drastically towards the cylinder boundary of 20 kpc. The flux decreases smoothly for $z<1$ kpc and drops quickly as $z$ approaches 4 kpc.

\subsection{\label{sec_CRDM}CReDM flux calculation}
The calculation process of the CReDM flux follows the steps articulated in previous works~\cite{CRDM_prl,CRDM_neutrino,CRDM_PRD,CRDM_TJ,CRDM_ZYF}. The DM particles with velocities about $ 10^{-3}\ c$ can be effectively treated as at rest when compared to the CREs. The kinetic energy transferred to a DM particle by an incident CRE particle with mass $m_e$ and kinetic energy $T_e$ is:
\begin{equation} \label{eq2}
\begin{aligned}
T_{\chi}&=T^{max}_{\chi}\frac{1-\rm cos{\mathrm{\theta}}}{2},\\
T_{\chi}^{max}&=\frac{T_e^2+2m_eT_e}{T_e+(m_e+m_{\chi})^2/(2m_{\chi})}.
\end{aligned}
\end{equation}

The $\mathrm{\theta}$ represents the scattering angle in the center-of-mass system. Inverting Eq.~\ref{eq2} indicates the minimal incident energy required to produce $T_{\chi}$:
\begin{equation} \label{invert}
  T_e^{min}=\left(\frac{T_{\chi}}{2}-m_e\right)\left(1\pm\sqrt{1+\frac{2T_{\chi}}{m_{\chi}}\frac{(m_e+m_{\chi})^2}{(2m_e-T_{\chi})^2}}\right) , 
\end{equation}
where $+(-)$ corresponds to the case $T_{\chi}>m_e$.

The total CReDM flux observed on the Earth is obtained by integrating over all CR nucleus species $i$ and energies $T_i$ along the line-of-sight (l.o.s.):
\begin{equation} \label{eq:crdm}
\begin{aligned}
\frac{d\Phi_{\chi}}{dT_{\chi}}&=\int\frac{d\Omega}{4\pi}\int_{l.o.s.} dl \int_{T_{e}^{min}}^{\infty}\frac{\rho_{\chi}(\mathbf{r})}{m_{\chi}}\frac{d\sigma_{\chi e}}{dT_{\chi}}\frac{d\Phi_e(\mathbf{r})}{dT_e}dT_e,
\end{aligned}
\end{equation}
where $\Omega$ is the solid angle and $T_{e}^{min}$ is evaluated by Eq.~\ref{invert}. The CRE flux, $d\Phi_e(\mathbf{r})/dT_e$ is related to the distance to the GC $\mathbf{r}$ and generated by the $\tt GALPROP$ code. For the DM density distribution $\rho_{\chi}(\mathbf{r})$, we adopt the Navarro-Frenk-White (NFW) profile~\cite{nfw}. The DM density reads:
\begin{equation} \label{density}
\begin{aligned}
\rho_{\chi}^{\rm NFW}(r)=\rho_{\chi}^{loc}\left(1+\frac{R_{\odot}}{R_S}\right)^2\left(\frac{r}{R_{\odot}}\right)^{-1}\left(1+\frac{r}{R_S}\right)^{-2},
\end{aligned}
\end{equation}
where $R_S = 20$ kpc, $\rho_{\chi}^{loc} = 0.3 \rm\ GeV\cdot cm^{-3}$.

The CRE flux and the DM density profile were regarded as homogeneous in many previous CReDM analyses~\cite{CRDM_neutrino,CRDM_PRD}. The local CRE flux and DM density near the Earth were used and the spatial integral was represented by the effective distance $D_{eff}$. The $D_{eff}$ values range from 1--10 kpc, becoming a major source of uncertainties. This work considers the spatial distribution of the CRE flux and the DM density profile instead of using the simplified $D_{eff}$, which gives more accurate CReDM fluxes. 

The $\chi$-$e$ elastic scattering cross section was assumed to be energy independent in previous studies~\cite{CRDM_PRD,CRDM_neutrino}, which reads:
\begin{equation} \label{independent}
\begin{aligned}
\frac{d\sigma_{\chi e}}{d T_{\chi}}=\frac{\Bar{\sigma}_{\chi e}}{T_{\chi}^{max}}.
\end{aligned}
\end{equation}

However, the cross section can depend on the kinetic energy regarding the type of coupling. We consider a fermionic DM particle, which interacts with electron through a scalar mediator $\phi$: $\mathscr{L}=g_{\chi \phi} \phi \bar{\chi} \chi+g_{e \phi} \phi \bar{e} e$. This assumption agrees with several leptophilic DM models. Following the treatment in previous study~\cite{energy_dependent}, the energy dependent cross section can be written as:
\begin{equation} \label{cross section}
\begin{aligned}
\frac{d\sigma_{\chi e}}{d T_{\chi}}=\bar{\sigma}_{\chi e}\frac{m_e^2m_{\chi}^2}{\mu_{\chi e}^2}\frac{F^2_{\rm DM}(q^2)}{[(m_{\chi }+m_e)^2+2m_{\chi}T_e]T_{\chi}^{max}},
\end{aligned}
\end{equation}
where $F^2_{\rm DM}(q^2)$ is the DM form factor with respect to the momentum transfer $q$ and the $\mu_{\chi e}$ is the reduced mass of DM and electron. The expression of the form factor depends on the form of the coupling. In this analysis, we have considered two cases: the case of the heavy mediator $F_{\rm DM}\propto 1$ and the case of the light mediator $F_{\rm DM}\propto1/q^2$. The energy dependent cross section in the heavy mediator case is:
\begin{equation} \label{heavy}
\begin{aligned}
\frac{d\sigma_{\chi e}}{d T_{\chi}}=\bar{\sigma}_{\chi e}\frac{m_{\chi}}{4\mu_{\chi e}^2}\frac{(2m_{\chi}+T_{\chi})(2m_e^2+m_{\chi}T_{\chi})}{[(m_{\chi }+m_e)^2+2m_{\chi}T_e]T_{\chi}^{max}}.
\end{aligned}
\end{equation}

The energy dependent cross section in the light mediator case is:
\begin{equation} \label{light}
\begin{aligned}
\frac{d\sigma_{\chi e}}{d T_{\chi}}=\bar{\sigma}_{\chi e}\frac{q^2_{ref}}{q^2}\frac{m_{\chi}}{4\mu_{\chi e}^2}\frac{(2m_{\chi}+T_{\chi})(2m_e^2+m_{\chi}T_{\chi})}{[(m_{\chi }+m_e)^2+2m_{\chi}T_e]T_{\chi}^{max}},
\end{aligned}
\end{equation}
where $q_{ref}=\alpha m_e$ is the reference momentum, $\alpha$ is the fine structure constant. 

\begin{figure}[!htbp]
\includegraphics[width=\linewidth]{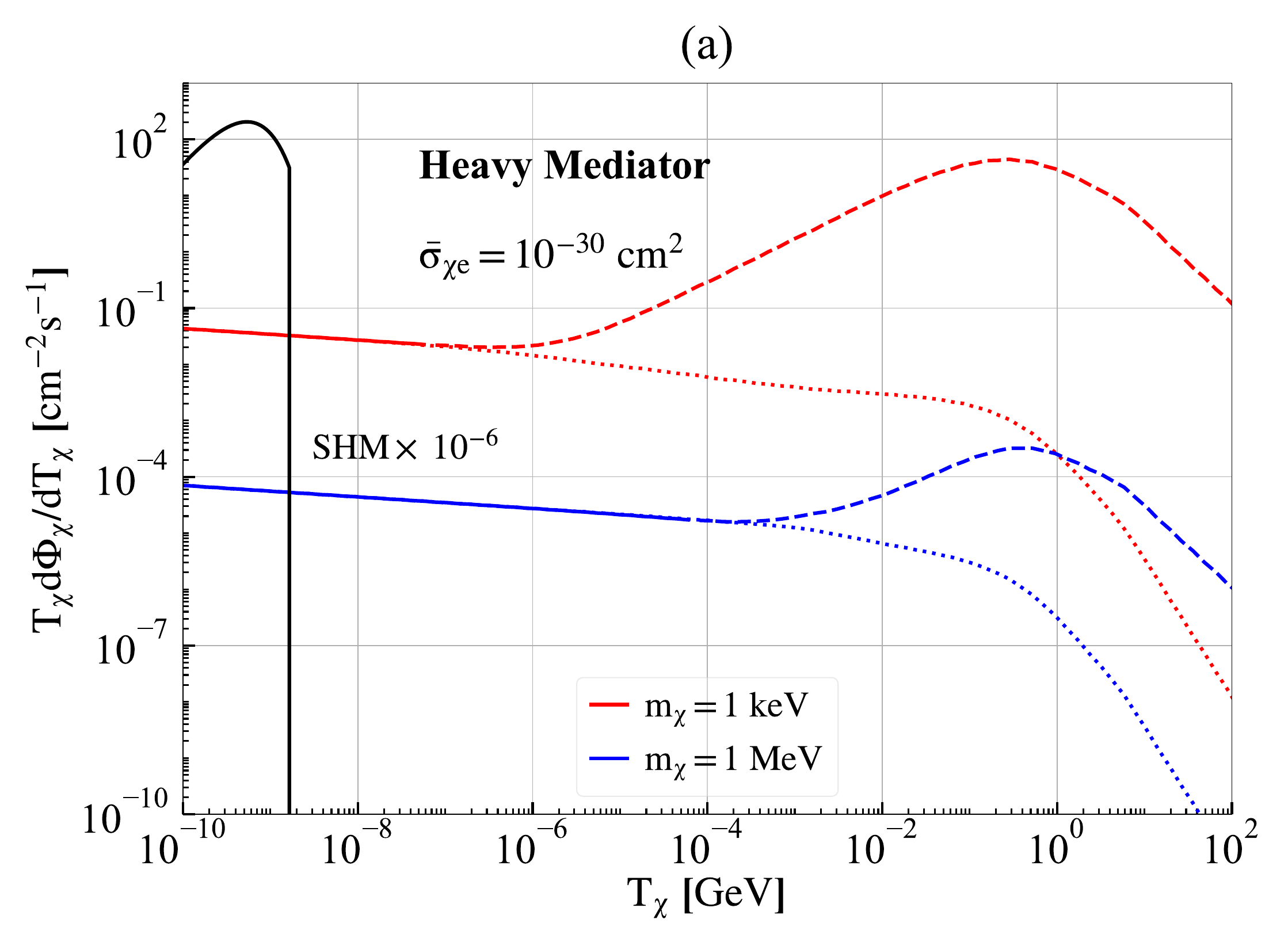}
\includegraphics[width=\linewidth]{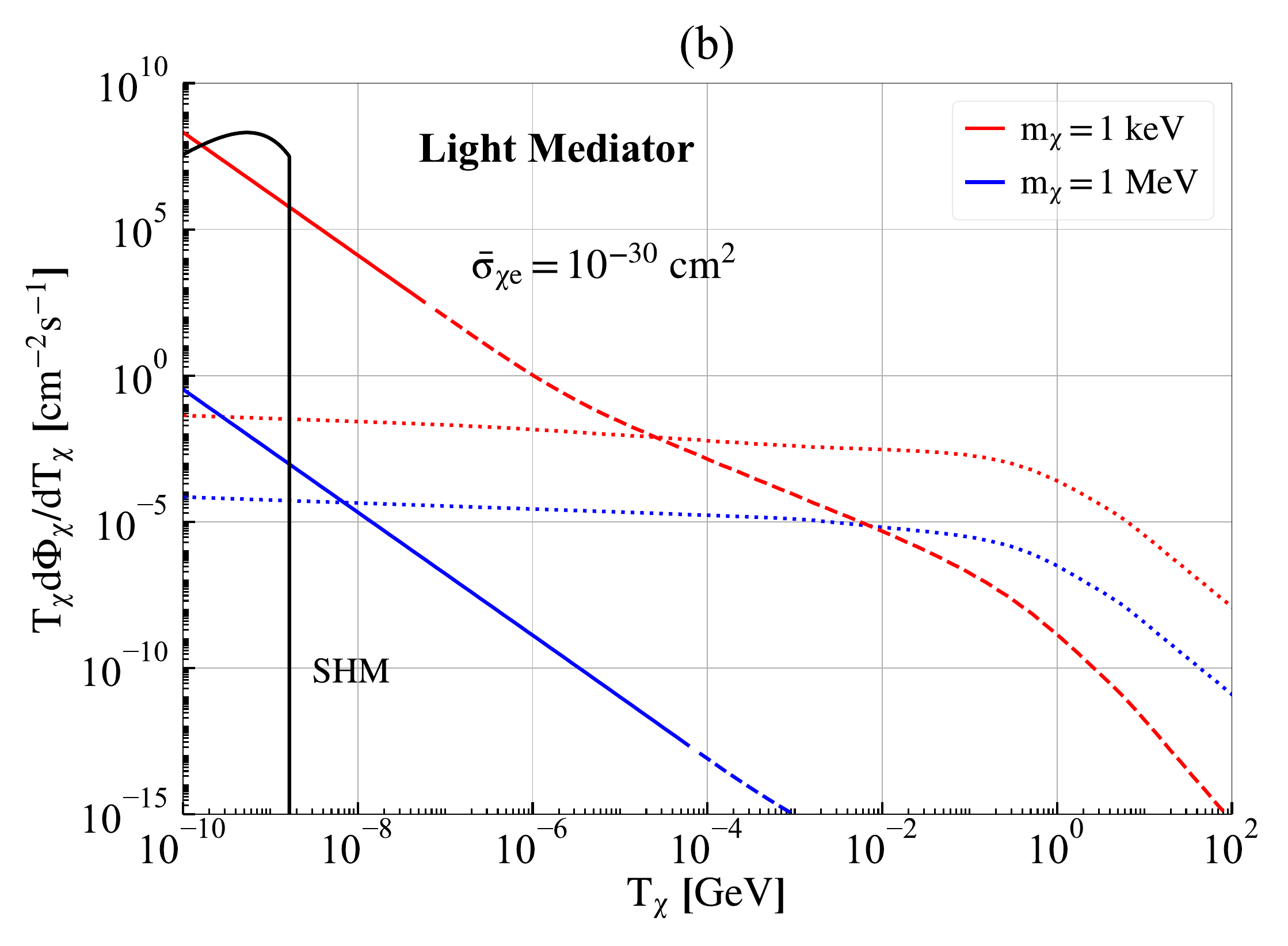}
\caption{The CReDM fluxes scaled by $T_{\chi}$ for different DM masses with $\bar{\sigma}_{\chi e} = 10^{-30}\ \rm cm^2$. The solid and dashed lines in upper and lower panel correspond to the heavy and light mediator cases, respectively. When the boosted DM velocity is more than $10^5\ \rm km/s$, the solid line is changed to dashed line. The dotted lines are calculated by using the energy independent cross section in Eq.~\ref{independent}. The DM flux under the SHM scenario with mass 1 MeV is plotted for comparison. The SHM flux in (a) is scaled by $10^{-6}$.}
\label{fig::crdm}
\end{figure}

Combining Eq.~\ref{eq:crdm}, Eq.~\ref{heavy} and Eq.~\ref{light}, the CReDM spectra in the space for both heavy and light mediator cases have been calculated and shown in Fig.~\ref{fig::crdm}. In the heavy mediator case, the CReDM fluxes are enhanced compared to the results from energy independent scenario in high energy regions for all DM masses. In the light mediator case, the fluxes are enhanced for small DM masses while suppressed for large DM masses. Compared with the SHM distribution, it can be seen that a small fraction of DM particles are boosted to very large kinetic energies. We assume that DM only couples with electron, namely $\sigma_{\chi N}=0$, so the attenuation effect of the rock over the underground laboratory is negligible~\cite{earthshielding4}. The CReDM fluxes calculated by Eq.~\ref{eq:crdm} can be regarded as the fluxes reaching the detectors underground.

\section{\label{sec3}Expected Spectra in Ge Detectors}
The calculation processes of the expected $\chi$-$e$ elastic scattering rates in the noble gases are well developed, while theoretical calculations of $\chi$-$e$ rates in semiconductors are more complicated. The atoms in the noble gases can be treated as isolated, and the wave functions and energy levels are well tabulated~\cite{noble_gas_table}, whereas the atom states are bound with the crystal environments for semiconductors. The $\chi$-$e$ transition rate per target mass in semiconductor detectors $R_{i\rightarrow f}$ reads~\cite{Rif1,Rif2,Rif3}:
\begin{equation}\label{eq:Rif}
\begin{aligned}
R_{i\rightarrow f}=&\frac{2\pi \bar{\sigma}_{\chi e}}{V\mu_{\chi e}^2 m_{\chi}}\frac{\rho_{\chi}^{loc}}{\rho_T}\\
&\sum_{i,f}\int \frac{d^3 q}{(2\pi)^3}\left(\frac{f_e}{f_e^0}\right)^2F^2_{\rm DM}g(\textit{\textbf{q}},\omega)|f_{i\rightarrow f} (\textit{\textbf{q}})|^2,
\end{aligned}
\end{equation}
\begin{equation}
\bar{\sigma}_{\chi e}=\frac{\mu_{\chi e}^2}{16\pi m_{\chi}^2 m_e^2}\overline{|\mathcal{M}(q_0)|^2},
\end{equation}
\begin{equation}
f_{i\rightarrow f}=\int d^3 x e^{i\textit{\textbf{q}} \cdot \textit{\textbf{x}}}\psi^*_f (\textit{\textbf{x}})\psi_i(\textit{\textbf{x}}) ,
\end{equation}
where $\rho_T$ is the target density, $V$ is the target volume, $\mu_{\chi e}$ is the DM electron reduced mass, $\textit{\textbf{q}}$ is the DM momentum, $f_{i\rightarrow f}$ is the crystal form factor depending on momentum transfer, $\bar{\sigma}_{\chi e}$ is the reference cross section for free electron scattering~\cite{Rif2}. The spin average matrix element $|\mathcal{M}(\textit{\textbf{q}})|^2$ can be factorized as $\mathcal{M}(\textit{\textbf{q}})=\mathcal{M}(q_0)(f_e/f_e^0)F_{\rm DM}$ for simple DM models, where the reference momentum transfer $q_0$ is taken as $\alpha m_e$ and $f_e/f_e^0$ is the screening factor. The $g(\textit{\textbf{q}},\omega)$ in Eq.~\ref{eq:Rif}is the velocity integral, $g (\textit{\textbf{q}}, \omega)=2 \pi\int d^{3} v f_{\chi} (\textit{\textbf{v}} ) \delta\left( \omega-\omega_{\bf q} \right)$, where $f_{\chi} (\textit{\textbf{v}})$ is the DM velocity distribution in the lab frame. 

Previously, CDEX has developed the velocity component analysis (VCA) method to calculate the expected spectra of DM particles with non-Maxwell-Boltzmann distribution~\cite{CDEX:SRDM}. The germanium detector responses to the CReDM are evaluated by the VCA method, which is developed based on the $\tt EXCEED$-$\tt DM$ package~\cite{exceed_dm}, replacing its default Maxwell-Boltzmann distribution with the CReDM velocity profiles calculated by Eq.~\ref{eq:crdm}.

However, the calculation process in the $\tt EXCEED$-$\tt DM$ package is non-relativistic, therefore only the CReDM with velocity lower than $v_{max} = 10^5\ \rm km/s$ is considered in the calculation of event rate. The DM particles with velocities below $10^5\ \rm km/s$ have a Lorentz factor lower than 1.06, and can be safely treated as non-relativistic. As shown in Fig.~\ref{fig::crdm}, this cutoff has a minor effect on the light mediator case, as more than $99.6\%$ of the total CReDM fluxes are in the non-relativistic region, while the heavy mediator case is greatly affected as most of the fluxes are in the relativistic region and thus are eliminated, leading to a very conservative result. The method for relativistic $\chi-e$ scattering is under development, which can give more accurate results, especially in the heavy mediator case.

\begin{figure}[!htbp]
\includegraphics[width=\linewidth]{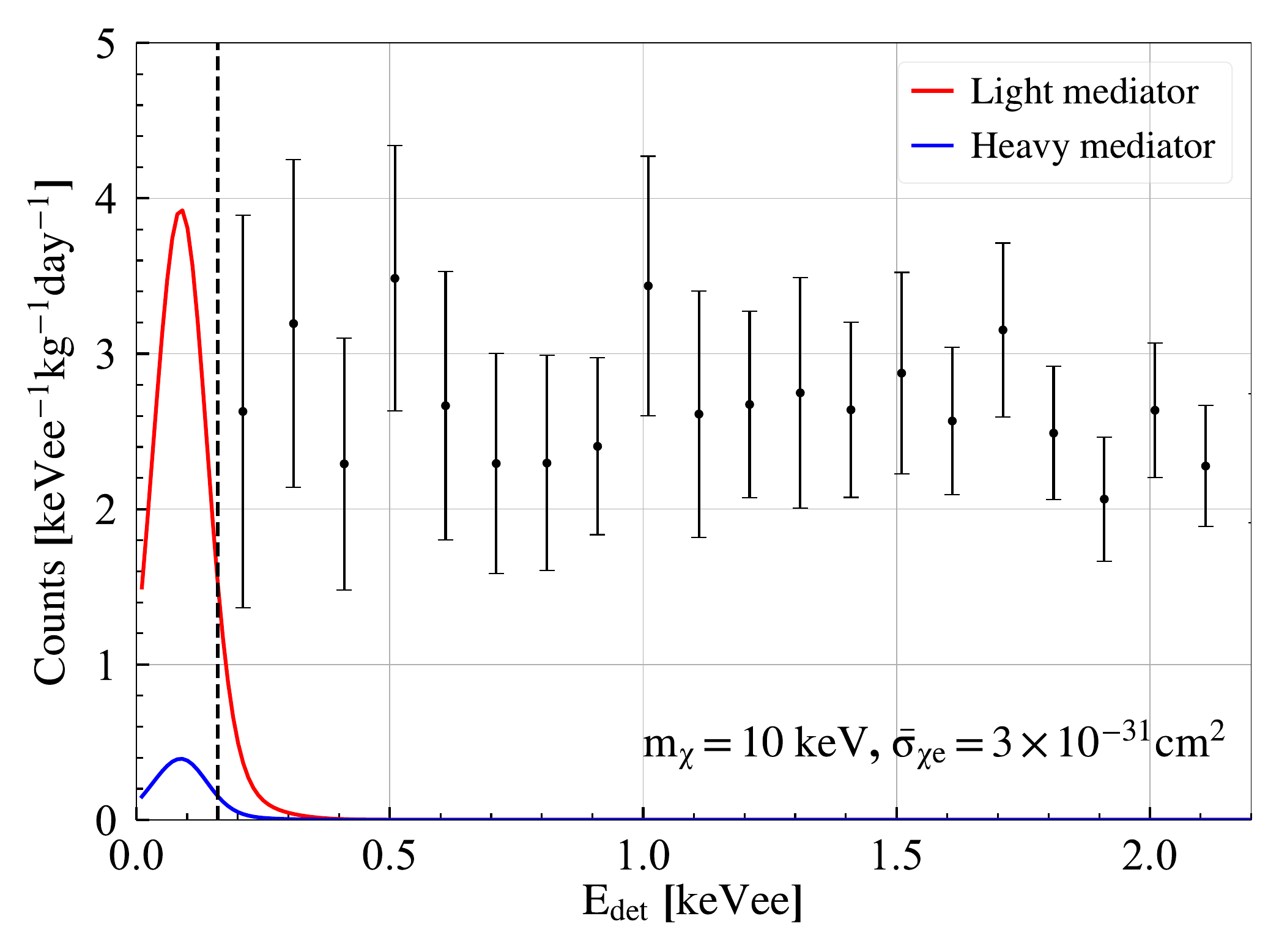}
\caption{The total expected rate of DM mass 10 keV and $\chi$-$e$ scattering cross section $10^{-31} \rm cm^2$. The red and blue lines correspond to the light and heavy mediator cases, respectively. The spectra are corrected with energy resolution. The standard deviation of the energy resolution is 35.8 + 16.6$\times E^{\frac {1}{2}}$ (eV), where $E$ is expressed in keV. The CDEX-10 experiment data after subtracting the contributions from cosmogenic radionuclides X-ray peaks are also shown for comparison. The CDEX-10 analysis threshold of 160 eVee is shown with a dotted line.}
\label{fig::spectrum}
\end{figure}

The expected spectra for both light and heavy mediator cases are shown in Fig.~\ref{fig::spectrum}. The event rates of the heavy mediator case are much lower than the rates of the light mediator case, mainly due to the velocity cutoff.

\section{\label{sec4}Exclusion results}
Data used in this CReDM analysis are from the CDEX-10 dataset, with a total exposure of 205.4 kg$\cdot$day~\cite{cdex_dark+pthoton}, after considering the dead time corrections and the fiducial mass. The data analysis follows the procedures described in our earlier works, including energy calibration, physics event selections, bulk-surface event discrimination and various efficiency corrections~\cite{cdex1b2018,cdex10_tech,cdex102018,cdex1b_am,cdex_dark+pthoton}. The physics analysis threshold is 160 eVee where the combined efficiency is 4.5\%. The characteristic K-shell X-ray peaks from internal cosmogenic radionuclides like $\rm ^{68}Ge$, $\rm ^{68}Ga$, $\rm ^{65}Zn$, $\rm ^{55}Fe$, $\rm ^{54}Mn$ and $\rm ^{49}V$ can be identified and their intensities are derived by the best fit of the spectrum~\cite{cdex102018}. The contributions of L- and M-shell X-ray peaks are derived from the corresponding K-shell line intensities~\cite{K-X-ray}. The residual CDEX-10 spectrum after subtracting the characteristic X-ray peaks from the cosmogenic radionuclides is used to give constraints on the CReDM.

A minimal-$\chi^2$ method is applied to calculate the exclusion region using the residual spectrum from 0.16--2.16 keVee, following the treatment in our previous works~\cite{cdex12014,cdex102018,cdex_dark+pthoton}. The $\chi^2$ value at certain DM mass $m_{\chi}$ and cross section $\bar{\sigma}_{\chi e}$ is defined by:
\begin{equation}
    \chi^2(m_{\chi},\bar{\sigma}_{\chi e}) = \sum_{i=0}^{N}\frac{[n_i-S_i(m_{\chi},\bar{\sigma}_{\chi e})-B_i]^2}{\sigma^2_{stat, i}+\sigma^2_{syst, i}},
\end{equation}
where $n_i$ is the measured event rate corresponding to the $i_{th}$ energy bin, and $S_i(m_{\chi},\bar{\sigma}_{\chi e})$ is the expected rate. The $\sigma_{stat}$ and $\sigma_{syst}$ corresponds to the statistical and systematical error, respectively. The background contribution at the $i_{th}$ energy bin is represented by $B_i$, which we assume to be a flat distribution; i.e., $B_i=B$. The best estimation of $\bar{\sigma}_{\chi e}$ at a given DM mass is obtained by minimizing the $\chi^2$ value. The upper limits at the 90\% confidence level (C.L.) are derived from the best estimation. 

\begin{figure}[!htbp]
\includegraphics[width=\linewidth]{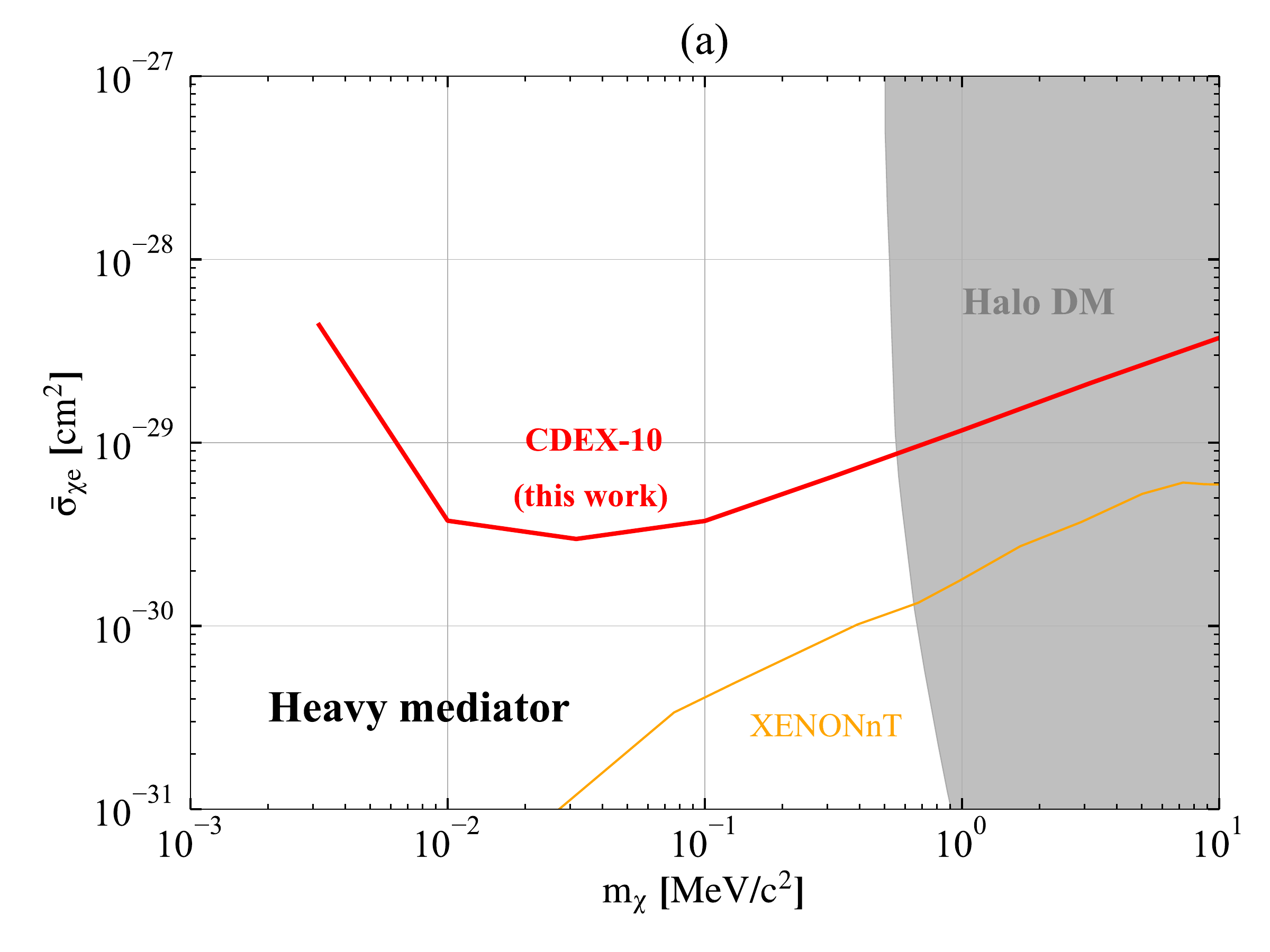}
\includegraphics[width=\linewidth]{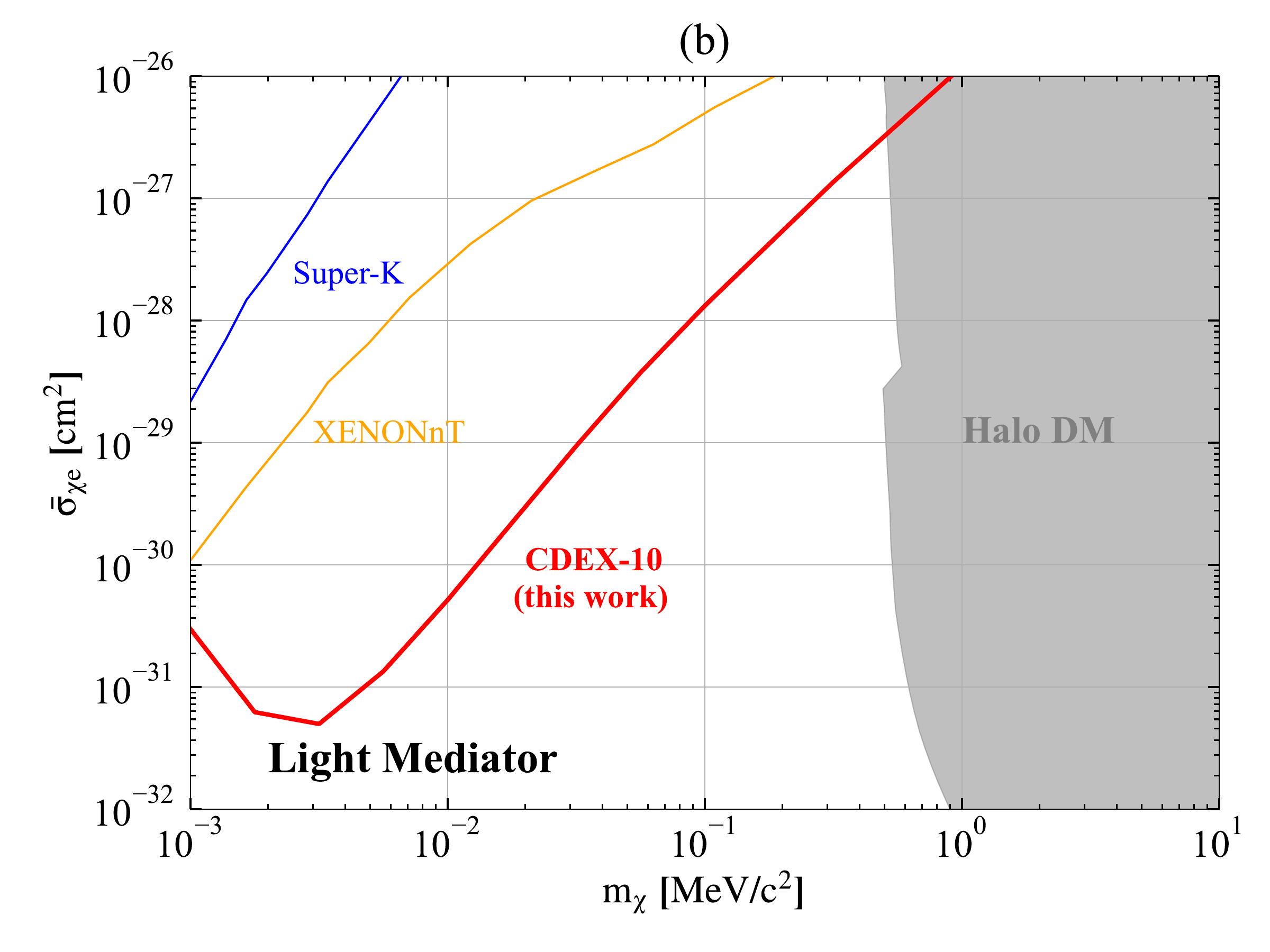}
\caption{The exclusion region derived from the CDEX-10 dataset is demonstrated in red line, with the CRE acceleration mechanism adopted. The upper and lower panel corresponds to the heavy and light mediator cases, respectively. The shaded area corresponds to the SHM constraints from current direct detection experiments~\cite{SENSEI,DAMIC}. Exclusion regions from the phenomenological CReDM analyses using XENONnT and Super-K data~\cite{energy_dependent} for both the heavy and light scalar mediator case are also shown for comparison.}
\label{fig::region}
\end{figure}

The exclusion lines calculated by the minimal-$\chi^2$ method using the CDEX-10 data for both the heavy and light mediator cases are shown in Fig.~\ref{fig::region}. The SHM constraints are expressed by the shaded area, the edge of which corresponds to the most stringent limits from the SENSEI~\cite{SENSEI} and the DAMIC~\cite{DAMIC} published results. The CReDM limits from phenomenological analyses using the XENONnT and Super-K data~\cite{energy_dependent} are also shown for comparison. The CDEX CReDM constraint exceeds the SHM limits under DM mass 0.6 MeV in the heavy mediator case. In the light mediator case, this work gives the most stringent limits for DM mass region from 1 keV to 0.5 MeV.

\newpage
\section{\label{sec5} Conclusions and discussions.}
Light dark matter particles boosted by CRE are detectable in the sub-GeV mass region. Novel computation techniques facilitate the CDEX to probe $\chi$-$e$ scattering in germanium detectors. This work sets exclusion limits using the CDEX-10 data under the CRE boosting scenario for both the heavy and light mediator cases. The result exceeds previous SHM limits for DM mass lighter than 0.6 MeV in the heavy mediator case and corresponds to the best sensitivity among all experiments from 1 keV to 0.5 MeV in the light mediator scenario.

The CRE spectra used for the CReDM calculation are generated by the $\tt GALPROP$ code package. The $\tt GALPROP$ code is also used to generate the CRE fluxes and spatial distributions in the whole Galaxy. The CRE spectra are extrapolated in the low energy range below 1 MeV. This treatment is supported by studies on the cosmic ray ionization of molecular clouds~\cite{molecular_cloud1,molecular_cloud2}, while some uncertainties still remain as no directly measured data can be used to examine this extrapolation.

Instead of adopting the energy independent assumption of the $\chi$-$e$ scattering cross section, we use the energy dependent cross section and investigate both the heavy and light mediator scenarios. The CReDM fluxes are enhanced in high kinetic energy range in the heavy mediator case. In the light mediator case, the fluxes are enhanced for small DM masses but suppressed for large DM masses.

We use the self-developed VCA method to evaluate the detector response of the CReDM. The calculation process of the expected spectra in germanium detector is non-relativistic. The input CReDM spectra need to be converted to velocity distributions and the high-speed parts of the spectra have to be truncated. This treatment makes our results rather conservative, especially in the heavy mediator case as the CReDM fluxes are very high in high energy range for heavy mediator couplings. Further advancement in the computation techniques can incorporate the relativistic effect and improve the results especially for the heavy mediator cases.

This work was supported by the National Key Research and Development Program of China (Grants No. 2023YFA1607101 and No. 2022YFA1605000) and the National Natural Science Foundation of China (Grants No. 12322511, No. 12175112, and No. 123B2087). We acknowledge the Center of High performance computing, Tsinghua University for providing the facility support. We would like to thank CJPL and its staff for hosting and supporting the CDEX project. CJPL is jointly operated by Tsinghua University and Yalong River Hydropower Development Company.

\bibliography{CReDM}

 \newcommand{\noop}[1]{}
\begin{thebibliography}{67}%
\makeatletter
\providecommand \@ifxundefined [1]{%
 \@ifx{#1\undefined}
}%
\providecommand \@ifnum [1]{%
 \ifnum #1\expandafter \@firstoftwo
 \else \expandafter \@secondoftwo
 \fi
}%
\providecommand \@ifx [1]{%
 \ifx #1\expandafter \@firstoftwo
 \else \expandafter \@secondoftwo
 \fi
}%
\providecommand \natexlab [1]{#1}%
\providecommand \enquote  [1]{``#1''}%
\providecommand \bibnamefont  [1]{#1}%
\providecommand \bibfnamefont [1]{#1}%
\providecommand \citenamefont [1]{#1}%
\providecommand \href@noop [0]{\@secondoftwo}%
\providecommand \href [0]{\begingroup \@sanitize@url \@href}%
\providecommand \@href[1]{\@@startlink{#1}\@@href}%
\providecommand \@@href[1]{\endgroup#1\@@endlink}%
\providecommand \@sanitize@url [0]{\catcode `\\12\catcode `\$12\catcode
  `\&12\catcode `\#12\catcode `\^12\catcode `\_12\catcode `\%12\relax}%
\providecommand \@@startlink[1]{}%
\providecommand \@@endlink[0]{}%
\providecommand \url  [0]{\begingroup\@sanitize@url \@url }%
\providecommand \@url [1]{\endgroup\@href {#1}{\urlprefix }}%
\providecommand \urlprefix  [0]{URL }%
\providecommand \Eprint [0]{\href }%
\providecommand \doibase [0]{http://dx.doi.org/}%
\providecommand \selectlanguage [0]{\@gobble}%
\providecommand \bibinfo  [0]{\@secondoftwo}%
\providecommand \bibfield  [0]{\@secondoftwo}%
\providecommand \translation [1]{[#1]}%
\providecommand \BibitemOpen [0]{}%
\providecommand \bibitemStop [0]{}%
\providecommand \bibitemNoStop [0]{.\EOS\space}%
\providecommand \EOS [0]{\spacefactor3000\relax}%
\providecommand \BibitemShut  [1]{\csname bibitem#1\endcsname}%
\let\auto@bib@innerbib\@empty
%</preamble>
\bibitem [{\citenamefont {Navas}\ \emph {et~al.}(2024)\citenamefont {Navas}
  \emph {et~al.}}]{RPP2024}%
  \BibitemOpen
  \bibfield  {author} {\bibinfo {author} {\bibfnamefont {S.}~\bibnamefont
  {Navas}} \emph {et~al.} (\bibinfo {collaboration} {Particle Data Group}),\
  }\href {https://link.aps.org/doi/10.1103/PhysRevD.110.030001} {\bibfield
  {journal} {\bibinfo  {journal} {Phys. Rev. D}\ }\textbf {\bibinfo {volume}
  {110}},\ \bibinfo {pages} {030001} (\bibinfo {year} {2024})}\BibitemShut
  {NoStop}%
\bibitem [{\citenamefont {Bertone}\ \emph {et~al.}(2005)\citenamefont
  {Bertone}, \citenamefont {Hooper},\ and\ \citenamefont
  {Silk}}]{bertone_particle_2005}%
  \BibitemOpen
  \bibfield  {author} {\bibinfo {author} {\bibfnamefont {G.}~\bibnamefont
  {Bertone}}, \bibinfo {author} {\bibfnamefont {D.}~\bibnamefont {Hooper}}, \
  and\ \bibinfo {author} {\bibfnamefont {J.}~\bibnamefont {Silk}},\ }\href
  {\doibase https://doi.org/10.1016/j.physrep.2004.08.031} {\bibfield
  {journal} {\bibinfo  {journal} {Phys. Rep.}\ }\textbf {\bibinfo {volume}
  {405}},\ \bibinfo {pages} {279} (\bibinfo {year} {2005})}\BibitemShut
  {NoStop}%
\bibitem [{\citenamefont {Aprile}\ \emph {et~al.}(2025)\citenamefont {Aprile}
  \emph {et~al.}}]{XENON:2025vwd}%
  \BibitemOpen
  \bibfield  {author} {\bibinfo {author} {\bibfnamefont {E.}~\bibnamefont
  {Aprile}} \emph {et~al.} (\bibinfo {collaboration} {XENON Collaboration}),\
  }\href {\doibase 10.1103/msw4-t342} {\bibfield  {journal} {\bibinfo
  {journal} {Phys. Rev. Lett.}\ }\textbf {\bibinfo {volume} {135}},\ \bibinfo
  {pages} {221003} (\bibinfo {year} {2025})}\BibitemShut {NoStop}%
\bibitem [{\citenamefont {Aalbers}\ \emph {et~al.}(2025)\citenamefont {Aalbers}
  \emph {et~al.}}]{LZ:2025iaw}%
  \BibitemOpen
  \bibfield  {author} {\bibinfo {author} {\bibfnamefont {J.}~\bibnamefont
  {Aalbers}} \emph {et~al.} (\bibinfo {collaboration} {LZ Collaboration}),\
  }\href {\doibase 10.1103/nr92-jvt3} {\bibfield  {journal} {\bibinfo
  {journal} {Phys. Rev. Lett.}\ }\textbf {\bibinfo {volume} {134}},\ \bibinfo
  {pages} {241801} (\bibinfo {year} {2025})}\BibitemShut {NoStop}%
\bibitem [{\citenamefont {Zhang}\ \emph {et~al.}(2025)\citenamefont {Zhang}
  \emph {et~al.}}]{PandaX:2025rrz}%
  \BibitemOpen
  \bibfield  {author} {\bibinfo {author} {\bibfnamefont {M.}~\bibnamefont
  {Zhang}} \emph {et~al.} (\bibinfo {collaboration} {PandaX Collaboration}),\
  }\href {\doibase 10.1103/rtnh-jn8s} {\bibfield  {journal} {\bibinfo
  {journal} {Phys. Rev. Lett.}\ }\textbf {\bibinfo {volume} {135}},\ \bibinfo
  {pages} {211001} (\bibinfo {year} {2025})}\BibitemShut {NoStop}%
\bibitem [{\citenamefont {Agnes}\ \emph {et~al.}(2024)\citenamefont {Agnes}
  \emph {et~al.}}]{DarkSide-50:2023fgf}%
  \BibitemOpen
  \bibfield  {author} {\bibinfo {author} {\bibfnamefont {P.}~\bibnamefont
  {Agnes}} \emph {et~al.} (\bibinfo {collaboration} {DarkSide-50
  Collaboration}),\ }\href {\doibase 10.1103/PhysRevD.110.102006} {\bibfield
  {journal} {\bibinfo  {journal} {Phys. Rev. D}\ }\textbf {\bibinfo {volume}
  {110}},\ \bibinfo {pages} {102006} (\bibinfo {year} {2024})}\BibitemShut
  {NoStop}%
\bibitem [{\citenamefont {Abdelhameed}\ \emph {et~al.}(2019)\citenamefont
  {Abdelhameed} \emph {et~al.}}]{cresst}%
  \BibitemOpen
  \bibfield  {author} {\bibinfo {author} {\bibfnamefont {A.~H.}\ \bibnamefont
  {Abdelhameed}} \emph {et~al.} (\bibinfo {collaboration} {CRESST
  Collaboration}),\ }\href {\doibase 10.1103/PhysRevD.100.102002} {\bibfield
  {journal} {\bibinfo  {journal} {Phys. Rev. D}\ }\textbf {\bibinfo {volume}
  {100}},\ \bibinfo {pages} {102002} (\bibinfo {year} {2019})}\BibitemShut
  {NoStop}%
\bibitem [{\citenamefont {Agnese}\ \emph {et~al.}(2018)\citenamefont {Agnese}
  \emph {et~al.}}]{cdmslite}%
  \BibitemOpen
  \bibfield  {author} {\bibinfo {author} {\bibfnamefont {R.}~\bibnamefont
  {Agnese}} \emph {et~al.} (\bibinfo {collaboration} {SuperCDMS
  Collaboration}),\ }\href {\doibase 10.1103/PhysRevD.97.022002} {\bibfield
  {journal} {\bibinfo  {journal} {Phys. Rev. D}\ }\textbf {\bibinfo {volume}
  {97}},\ \bibinfo {pages} {022002} (\bibinfo {year} {2018})}\BibitemShut
  {NoStop}%
\bibitem [{\citenamefont {Aalseth}\ \emph {et~al.}(2013)\citenamefont {Aalseth}
  \emph {et~al.}}]{cogent2013}%
  \BibitemOpen
  \bibfield  {author} {\bibinfo {author} {\bibfnamefont {C.~E.}\ \bibnamefont
  {Aalseth}} \emph {et~al.} (\bibinfo {collaboration} {CoGeNT Collaboration}),\
  }\href {\doibase 10.1103/PhysRevD.88.012002} {\bibfield  {journal} {\bibinfo
  {journal} {Phys. Rev. D}\ }\textbf {\bibinfo {volume} {88}},\ \bibinfo
  {pages} {012002} (\bibinfo {year} {2013})}\BibitemShut {NoStop}%
\bibitem [{\citenamefont {Liu}\ \emph {et~al.}(2014)\citenamefont {Liu} \emph
  {et~al.}}]{cdex0}%
  \BibitemOpen
  \bibfield  {author} {\bibinfo {author} {\bibfnamefont {S.~K.}\ \bibnamefont
  {Liu}} \emph {et~al.} (\bibinfo {collaboration} {CDEX Collaboration}),\
  }\href {\doibase 10.1103/PhysRevD.90.032003} {\bibfield  {journal} {\bibinfo
  {journal} {Phys. Rev. D}\ }\textbf {\bibinfo {volume} {90}},\ \bibinfo
  {pages} {032003} (\bibinfo {year} {2014})}\BibitemShut {NoStop}%
\bibitem [{\citenamefont {Zhao}\ \emph {et~al.}(2013)\citenamefont {Zhao} \emph
  {et~al.}}]{cdex1}%
  \BibitemOpen
  \bibfield  {author} {\bibinfo {author} {\bibfnamefont {W.}~\bibnamefont
  {Zhao}} \emph {et~al.} (\bibinfo {collaboration} {CDEX Collaboration}),\
  }\href {\doibase 10.1103/PhysRevD.88.052004} {\bibfield  {journal} {\bibinfo
  {journal} {Phys. Rev. D}\ }\textbf {\bibinfo {volume} {88}},\ \bibinfo
  {pages} {052004} (\bibinfo {year} {2013})}\BibitemShut {NoStop}%
\bibitem [{\citenamefont {Yue}\ \emph {et~al.}(2014)\citenamefont {Yue} \emph
  {et~al.}}]{cdex12014}%
  \BibitemOpen
  \bibfield  {author} {\bibinfo {author} {\bibfnamefont {Q.}~\bibnamefont
  {Yue}} \emph {et~al.} (\bibinfo {collaboration} {CDEX Collaboration}),\
  }\href {\doibase 10.1103/PhysRevD.90.091701} {\bibfield  {journal} {\bibinfo
  {journal} {Phys. Rev. D}\ }\textbf {\bibinfo {volume} {90}},\ \bibinfo
  {pages} {091701} (\bibinfo {year} {2014})}\BibitemShut {NoStop}%
\bibitem [{\citenamefont {Zhao}\ \emph {et~al.}(2016)\citenamefont {Zhao} \emph
  {et~al.}}]{cdex12016}%
  \BibitemOpen
  \bibfield  {author} {\bibinfo {author} {\bibfnamefont {W.}~\bibnamefont
  {Zhao}} \emph {et~al.} (\bibinfo {collaboration} {CDEX Collaboration}),\
  }\href {\doibase 10.1103/PhysRevD.93.092003} {\bibfield  {journal} {\bibinfo
  {journal} {Phys. Rev. D}\ }\textbf {\bibinfo {volume} {93}},\ \bibinfo
  {pages} {092003} (\bibinfo {year} {2016})}\BibitemShut {NoStop}%
\bibitem [{\citenamefont {Yang}\ \emph {et~al.}(2018)\citenamefont {Yang} \emph
  {et~al.}}]{cdex1b2018}%
  \BibitemOpen
  \bibfield  {author} {\bibinfo {author} {\bibfnamefont {L.~T.}\ \bibnamefont
  {Yang}} \emph {et~al.} (\bibinfo {collaboration} {CDEX Collaboration}),\
  }\href {\doibase 10.1088/1674-1137/42/2/023002} {\bibfield  {journal}
  {\bibinfo  {journal} {Chin. Phys. C}\ }\textbf {\bibinfo {volume} {42}},\
  \bibinfo {eid} {023002} (\bibinfo {year} {2018})}\BibitemShut {NoStop}%
\bibitem [{\citenamefont {Jiang}\ \emph {et~al.}(2018)\citenamefont {Jiang}
  \emph {et~al.}}]{cdex102018}%
  \BibitemOpen
  \bibfield  {author} {\bibinfo {author} {\bibfnamefont {H.}~\bibnamefont
  {Jiang}} \emph {et~al.} (\bibinfo {collaboration} {CDEX Collaboration}),\
  }\href {\doibase 10.1103/PhysRevLett.120.241301} {\bibfield  {journal}
  {\bibinfo  {journal} {Phys. Rev. Lett.}\ }\textbf {\bibinfo {volume} {120}},\
  \bibinfo {pages} {241301} (\bibinfo {year} {2018})}\BibitemShut {NoStop}%
\bibitem [{\citenamefont {Jiang}\ \emph {et~al.}(2019)\citenamefont {Jiang}
  \emph {et~al.}}]{cdex10_tech}%
  \BibitemOpen
  \bibfield  {author} {\bibinfo {author} {\bibfnamefont {H.}~\bibnamefont
  {Jiang}} \emph {et~al.} (\bibinfo {collaboration} {CDEX Collaboration}),\
  }\href {\doibase 10.1007/s11433-018-8001-3} {\bibfield  {journal} {\bibinfo
  {journal} {Sci. China Phys. Mech. Astron.}\ }\textbf {\bibinfo {volume}
  {62}},\ \bibinfo {pages} {31012} (\bibinfo {year} {2019})}\BibitemShut
  {NoStop}%
\bibitem [{\citenamefont {Yang}\ \emph {et~al.}(2019)\citenamefont {Yang} \emph
  {et~al.}}]{cdex1b_am}%
  \BibitemOpen
  \bibfield  {author} {\bibinfo {author} {\bibfnamefont {L.~T.}\ \bibnamefont
  {Yang}} \emph {et~al.} (\bibinfo {collaboration} {CDEX Collaboration}),\
  }\href {\doibase 10.1103/PhysRevLett.123.221301} {\bibfield  {journal}
  {\bibinfo  {journal} {Phys. Rev. Lett.}\ }\textbf {\bibinfo {volume} {123}},\
  \bibinfo {pages} {221301} (\bibinfo {year} {2019})}\BibitemShut {NoStop}%
\bibitem [{\citenamefont {Wang}\ \emph {et~al.}(2021)\citenamefont {Wang} \emph
  {et~al.}}]{cdex10_eft}%
  \BibitemOpen
  \bibfield  {author} {\bibinfo {author} {\bibfnamefont {Y.}~\bibnamefont
  {Wang}} \emph {et~al.} (\bibinfo {collaboration} {CDEX Collaboration}),\
  }\href {\doibase 10.1007/s11433-020-1666-8} {\bibfield  {journal} {\bibinfo
  {journal} {Sci. China-Phys. Mech. Astron.}\ }\textbf {\bibinfo {volume}
  {64}},\ \bibinfo {pages} {281011} (\bibinfo {year} {2021})}\BibitemShut
  {NoStop}%
\bibitem [{\citenamefont {Zhang}\ \emph {et~al.}(2022)\citenamefont {Zhang}
  \emph {et~al.}}]{CDEX_electron}%
  \BibitemOpen
  \bibfield  {author} {\bibinfo {author} {\bibfnamefont {Z.~Y.}\ \bibnamefont
  {Zhang}} \emph {et~al.} (\bibinfo {collaboration} {CDEX Collaboration}),\
  }\href {\doibase 10.1103/PhysRevLett.129.221301} {\bibfield  {journal}
  {\bibinfo  {journal} {Phys. Rev. Lett.}\ }\textbf {\bibinfo {volume} {129}},\
  \bibinfo {pages} {221301} (\bibinfo {year} {2022})}\BibitemShut {NoStop}%
\bibitem [{\citenamefont {Zhang}\ \emph {et~al.}(2023)\citenamefont {Zhang}
  \emph {et~al.}}]{CDEX_black_hole}%
  \BibitemOpen
  \bibfield  {author} {\bibinfo {author} {\bibfnamefont {Z.~H.}\ \bibnamefont
  {Zhang}} \emph {et~al.} (\bibinfo {collaboration} {CDEX Collaboration}),\
  }\href {\doibase 10.1103/PhysRevD.108.052006} {\bibfield  {journal} {\bibinfo
   {journal} {Phys. Rev. D}\ }\textbf {\bibinfo {volume} {108}},\ \bibinfo
  {pages} {052006} (\bibinfo {year} {2023})}\BibitemShut {NoStop}%
\bibitem [{\citenamefont {Geng}\ \emph {et~al.}(2023)\citenamefont {Geng} \emph
  {et~al.}}]{CDEX_neutrino}%
  \BibitemOpen
  \bibfield  {author} {\bibinfo {author} {\bibfnamefont {X.~P.}\ \bibnamefont
  {Geng}} \emph {et~al.} (\bibinfo {collaboration} {CDEX Collaboration}),\
  }\href {\doibase 10.1103/PhysRevD.107.112002} {\bibfield  {journal} {\bibinfo
   {journal} {Phys. Rev. D}\ }\textbf {\bibinfo {volume} {107}},\ \bibinfo
  {pages} {112002} (\bibinfo {year} {2023})}\BibitemShut {NoStop}%
\bibitem [{\citenamefont {Zhang}\ \emph
  {et~al.}(2024{\natexlab{a}})\citenamefont {Zhang} \emph
  {et~al.}}]{CDEX_PBHvDM}%
  \BibitemOpen
  \bibfield  {author} {\bibinfo {author} {\bibfnamefont {Z.~H.}\ \bibnamefont
  {Zhang}} \emph {et~al.} (\bibinfo {collaboration} {CDEX Collaboration}),\
  }\href {\doibase 10.1007/s11433-024-2446-2} {\bibfield  {journal} {\bibinfo
  {journal} {Sci. China Phys. Mech. Astron.}\ }\textbf {\bibinfo {volume}
  {67}},\ \bibinfo {pages} {101011} (\bibinfo {year}
  {2024}{\natexlab{a}})}\BibitemShut {NoStop}%
\bibitem [{\citenamefont {Zhang}\ \emph
  {et~al.}(2024{\natexlab{b}})\citenamefont {Zhang} \emph
  {et~al.}}]{CDEX:SRDM}%
  \BibitemOpen
  \bibfield  {author} {\bibinfo {author} {\bibfnamefont {Z.~Y.}\ \bibnamefont
  {Zhang}} \emph {et~al.} (\bibinfo {collaboration} {CDEX Collaboration}),\
  }\href {\doibase 10.1103/PhysRevLett.132.171001} {\bibfield  {journal}
  {\bibinfo  {journal} {Phys. Rev. Lett.}\ }\textbf {\bibinfo {volume} {132}},\
  \bibinfo {pages} {171001} (\bibinfo {year} {2024}{\natexlab{b}})}\BibitemShut
  {NoStop}%
\bibitem [{\citenamefont {Wang}\ \emph {et~al.}(2025)\citenamefont {Wang} \emph
  {et~al.}}]{CDEX:supernova}%
  \BibitemOpen
  \bibfield  {author} {\bibinfo {author} {\bibfnamefont {J.~Z.}\ \bibnamefont
  {Wang}} \emph {et~al.} (\bibinfo {collaboration} {CDEX Collaboration}),\
  }\href {\doibase 10.1103/kxmk-7zfk} {\bibfield  {journal} {\bibinfo
  {journal} {Phys. Rev. D}\ }\textbf {\bibinfo {volume} {112}},\ \bibinfo
  {pages} {092011} (\bibinfo {year} {2025})}\BibitemShut {NoStop}%
\bibitem [{\citenamefont {Bloch}\ \emph {et~al.}(2025)\citenamefont {Bloch}
  \emph {et~al.}}]{SENSEI:2024yyt}%
  \BibitemOpen
  \bibfield  {author} {\bibinfo {author} {\bibfnamefont {I.~M.}\ \bibnamefont
  {Bloch}} \emph {et~al.} (\bibinfo {collaboration} {SENSEI Collaboration}),\
  }\href {\doibase 10.1103/PhysRevLett.134.161002} {\bibfield  {journal}
  {\bibinfo  {journal} {Phys. Rev. Lett.}\ }\textbf {\bibinfo {volume} {134}},\
  \bibinfo {pages} {161002} (\bibinfo {year} {2025})}\BibitemShut {NoStop}%
\bibitem [{\citenamefont {Aguilar-Arevalo}\ \emph {et~al.}(2024)\citenamefont
  {Aguilar-Arevalo} \emph {et~al.}}]{DAMIC:2023rcc}%
  \BibitemOpen
  \bibfield  {author} {\bibinfo {author} {\bibfnamefont {A.}~\bibnamefont
  {Aguilar-Arevalo}} \emph {et~al.} (\bibinfo {collaboration} {SENSEI, DAMIC-M,
  DAMIC Collaboration}),\ }\href {\doibase 10.1103/PhysRevD.109.062007}
  {\bibfield  {journal} {\bibinfo  {journal} {Phys. Rev. D}\ }\textbf {\bibinfo
  {volume} {109}},\ \bibinfo {pages} {062007} (\bibinfo {year}
  {2024})}\BibitemShut {NoStop}%
\bibitem [{\citenamefont {Arnaud}\ \emph {et~al.}(2020)\citenamefont {Arnaud}
  \emph {et~al.}}]{EDELWEISS_electron}%
  \BibitemOpen
  \bibfield  {author} {\bibinfo {author} {\bibfnamefont {Q.}~\bibnamefont
  {Arnaud}} \emph {et~al.} (\bibinfo {collaboration} {EDELWEISS
  Collaboration}),\ }\href {\doibase 10.1103/PhysRevLett.125.141301} {\bibfield
   {journal} {\bibinfo  {journal} {Phys. Rev. Lett.}\ }\textbf {\bibinfo
  {volume} {125}},\ \bibinfo {pages} {141301} (\bibinfo {year}
  {2020})}\BibitemShut {NoStop}%
\bibitem [{\citenamefont {Amaral}\ \emph {et~al.}(2020)\citenamefont {Amaral}
  \emph {et~al.}}]{supercdms}%
  \BibitemOpen
  \bibfield  {author} {\bibinfo {author} {\bibfnamefont {D.~W.}\ \bibnamefont
  {Amaral}} \emph {et~al.},\ }\href {\doibase 10.1103/PhysRevD.102.091101}
  {\bibfield  {journal} {\bibinfo  {journal} {Phys. Rev. D}\ }\textbf {\bibinfo
  {volume} {102}},\ \bibinfo {pages} {091101} (\bibinfo {year}
  {2020})}\BibitemShut {NoStop}%
\bibitem [{\citenamefont {Aprile}\ \emph {et~al.}(2019)\citenamefont {Aprile}
  \emph {et~al.}}]{XENON_electron}%
  \BibitemOpen
  \bibfield  {author} {\bibinfo {author} {\bibfnamefont {E.}~\bibnamefont
  {Aprile}} \emph {et~al.} (\bibinfo {collaboration} {XENON Collaboration}),\
  }\href {\doibase 10.1103/PhysRevLett.123.251801} {\bibfield  {journal}
  {\bibinfo  {journal} {Phys. Rev. Lett.}\ }\textbf {\bibinfo {volume} {123}},\
  \bibinfo {pages} {251801} (\bibinfo {year} {2019})}\BibitemShut {NoStop}%
\bibitem [{\citenamefont {Cheng}\ \emph {et~al.}(2021)\citenamefont {Cheng}
  \emph {et~al.}}]{pandax_electron}%
  \BibitemOpen
  \bibfield  {author} {\bibinfo {author} {\bibfnamefont {C.}~\bibnamefont
  {Cheng}} \emph {et~al.} (\bibinfo {collaboration} {PandaX-II
  Collaboration}),\ }\href {\doibase 10.1103/PhysRevLett.126.211803} {\bibfield
   {journal} {\bibinfo  {journal} {Phys. Rev. Lett.}\ }\textbf {\bibinfo
  {volume} {126}},\ \bibinfo {pages} {211803} (\bibinfo {year}
  {2021})}\BibitemShut {NoStop}%
\bibitem [{\citenamefont {Bringmann}\ and\ \citenamefont
  {Pospelov}(2019)}]{CRDM_prl}%
  \BibitemOpen
  \bibfield  {author} {\bibinfo {author} {\bibfnamefont {T.}~\bibnamefont
  {Bringmann}}\ and\ \bibinfo {author} {\bibfnamefont {M.}~\bibnamefont
  {Pospelov}},\ }\href {\doibase 10.1103/PhysRevLett.122.171801} {\bibfield
  {journal} {\bibinfo  {journal} {Phys. Rev. Lett.}\ }\textbf {\bibinfo
  {volume} {122}},\ \bibinfo {pages} {171801} (\bibinfo {year}
  {2019})}\BibitemShut {NoStop}%
\bibitem [{\citenamefont {Lei}\ \emph {et~al.}(2022)\citenamefont {Lei},
  \citenamefont {Tang},\ and\ \citenamefont {Zhang}}]{CRDM_TJ}%
  \BibitemOpen
  \bibfield  {author} {\bibinfo {author} {\bibfnamefont {Z.~H.}\ \bibnamefont
  {Lei}}, \bibinfo {author} {\bibfnamefont {J.}~\bibnamefont {Tang}}, \ and\
  \bibinfo {author} {\bibfnamefont {B.~L.}\ \bibnamefont {Zhang}},\ }\href
  {\doibase 10.1088/1674-1137/ac68da} {\bibfield  {journal} {\bibinfo
  {journal} {Chin. Phys. C}\ }\textbf {\bibinfo {volume} {46}},\ \bibinfo
  {pages} {085103} (\bibinfo {year} {2022})}\BibitemShut {NoStop}%
\bibitem [{\citenamefont {Xia}\ \emph {et~al.}(2021)\citenamefont {Xia},
  \citenamefont {Xu},\ and\ \citenamefont {Zhou}}]{CRDM_ZYF}%
  \BibitemOpen
  \bibfield  {author} {\bibinfo {author} {\bibfnamefont {C.}~\bibnamefont
  {Xia}}, \bibinfo {author} {\bibfnamefont {Y.~H.}\ \bibnamefont {Xu}}, \ and\
  \bibinfo {author} {\bibfnamefont {Y.~F.}\ \bibnamefont {Zhou}},\ }\href
  {\doibase https://doi.org/10.1016/j.nuclphysb.2021.115470} {\bibfield
  {journal} {\bibinfo  {journal} {Nucl. Phys. B}\ }\textbf {\bibinfo {volume}
  {969}},\ \bibinfo {pages} {115470} (\bibinfo {year} {2021})}\BibitemShut
  {NoStop}%
\bibitem [{\citenamefont {Cappiello}\ and\ \citenamefont
  {Beacom}(2019)}]{CRDM_PRD}%
  \BibitemOpen
  \bibfield  {author} {\bibinfo {author} {\bibfnamefont {C.~V.}\ \bibnamefont
  {Cappiello}}\ and\ \bibinfo {author} {\bibfnamefont {J.~F.}\ \bibnamefont
  {Beacom}},\ }\href {\doibase 10.1103/PhysRevD.100.103011} {\bibfield
  {journal} {\bibinfo  {journal} {Phys. Rev. D}\ }\textbf {\bibinfo {volume}
  {100}},\ \bibinfo {pages} {103011} (\bibinfo {year} {2019})}\BibitemShut
  {NoStop}%
\bibitem [{\citenamefont {Ema}\ \emph {et~al.}(2019)\citenamefont {Ema},
  \citenamefont {Sala},\ and\ \citenamefont {Sato}}]{CRDM_neutrino}%
  \BibitemOpen
  \bibfield  {author} {\bibinfo {author} {\bibfnamefont {Y.}~\bibnamefont
  {Ema}}, \bibinfo {author} {\bibfnamefont {F.}~\bibnamefont {Sala}}, \ and\
  \bibinfo {author} {\bibfnamefont {R.}~\bibnamefont {Sato}},\ }\href {\doibase
  10.1103/PhysRevLett.122.181802} {\bibfield  {journal} {\bibinfo  {journal}
  {Phys. Rev. Lett.}\ }\textbf {\bibinfo {volume} {122}},\ \bibinfo {pages}
  {181802} (\bibinfo {year} {2019})}\BibitemShut {NoStop}%
\bibitem [{\citenamefont {Dent}\ \emph {et~al.}(2020)\citenamefont {Dent} \emph
  {et~al.}}]{CRDM_prd2}%
  \BibitemOpen
  \bibfield  {author} {\bibinfo {author} {\bibfnamefont {J.~B.}\ \bibnamefont
  {Dent}} \emph {et~al.},\ }\href {\doibase 10.1103/PhysRevD.101.116007}
  {\bibfield  {journal} {\bibinfo  {journal} {Phys. Rev. D}\ }\textbf {\bibinfo
  {volume} {101}},\ \bibinfo {pages} {116007} (\bibinfo {year}
  {2020})}\BibitemShut {NoStop}%
\bibitem [{\citenamefont {Cui}\ \emph {et~al.}(2022)\citenamefont {Cui} \emph
  {et~al.}}]{PandaX-CRDM}%
  \BibitemOpen
  \bibfield  {author} {\bibinfo {author} {\bibfnamefont {X.}~\bibnamefont
  {Cui}} \emph {et~al.} (\bibinfo {collaboration} {PandaX-II Collaboration}),\
  }\href {\doibase 10.1103/PhysRevLett.128.171801} {\bibfield  {journal}
  {\bibinfo  {journal} {Phys. Rev. Lett.}\ }\textbf {\bibinfo {volume} {128}},\
  \bibinfo {pages} {171801} (\bibinfo {year} {2022})}\BibitemShut {NoStop}%
\bibitem [{\citenamefont {Xu}\ \emph {et~al.}(2022)\citenamefont {Xu} \emph
  {et~al.}}]{CDEX_CRDM}%
  \BibitemOpen
  \bibfield  {author} {\bibinfo {author} {\bibfnamefont {R.}~\bibnamefont {Xu}}
  \emph {et~al.} (\bibinfo {collaboration} {CDEX Collaboration}),\ }\href
  {\doibase 10.1103/PhysRevD.106.052008} {\bibfield  {journal} {\bibinfo
  {journal} {Phys. Rev. D}\ }\textbf {\bibinfo {volume} {106}},\ \bibinfo
  {pages} {052008} (\bibinfo {year} {2022})}\BibitemShut {NoStop}%
\bibitem [{\citenamefont {Xia}\ \emph {et~al.}(2023)\citenamefont {Xia},
  \citenamefont {Xu},\ and\ \citenamefont {Zhou}}]{aziumth_assymetry}%
  \BibitemOpen
  \bibfield  {author} {\bibinfo {author} {\bibfnamefont {C.}~\bibnamefont
  {Xia}}, \bibinfo {author} {\bibfnamefont {Y.-H.}\ \bibnamefont {Xu}}, \ and\
  \bibinfo {author} {\bibfnamefont {Y.-F.}\ \bibnamefont {Zhou}},\ }\href
  {\doibase 10.1103/PhysRevD.107.055012} {\bibfield  {journal} {\bibinfo
  {journal} {Phys. Rev. D}\ }\textbf {\bibinfo {volume} {107}},\ \bibinfo
  {pages} {055012} (\bibinfo {year} {2023})}\BibitemShut {NoStop}%
\bibitem [{\citenamefont {Adhikari}\ \emph {et~al.}(2023)\citenamefont
  {Adhikari} \emph {et~al.}}]{COSINE-100:2023tcq}%
  \BibitemOpen
  \bibfield  {author} {\bibinfo {author} {\bibfnamefont {G.}~\bibnamefont
  {Adhikari}} \emph {et~al.} (\bibinfo {collaboration} {COSINE-100
  Collaboration}),\ }\href {\doibase 10.1103/PhysRevLett.131.201802} {\bibfield
   {journal} {\bibinfo  {journal} {Phys. Rev. Lett.}\ }\textbf {\bibinfo
  {volume} {131}},\ \bibinfo {pages} {201802} (\bibinfo {year}
  {2023})}\BibitemShut {NoStop}%
\bibitem [{\citenamefont {Cappiello}\ \emph {et~al.}(2024)\citenamefont
  {Cappiello}, \citenamefont {Liu}, \citenamefont {Mohlabeng},\ and\
  \citenamefont {Vincent}}]{IceCube:BDM}%
  \BibitemOpen
  \bibfield  {author} {\bibinfo {author} {\bibfnamefont {C.~V.}\ \bibnamefont
  {Cappiello}}, \bibinfo {author} {\bibfnamefont {Q.}~\bibnamefont {Liu}},
  \bibinfo {author} {\bibfnamefont {G.}~\bibnamefont {Mohlabeng}}, \ and\
  \bibinfo {author} {\bibfnamefont {A.~C.}\ \bibnamefont {Vincent}},\ }\href
  {\doibase 10.1103/PhysRevD.110.095031} {\bibfield  {journal} {\bibinfo
  {journal} {Phys. Rev. D}\ }\textbf {\bibinfo {volume} {110}},\ \bibinfo
  {pages} {095031} (\bibinfo {year} {2024})}\BibitemShut {NoStop}%
\bibitem [{\citenamefont {Soma}\ \emph {et~al.}(2016)\citenamefont {Soma} \emph
  {et~al.}}]{soma2016}%
  \BibitemOpen
  \bibfield  {author} {\bibinfo {author} {\bibfnamefont {A.}~\bibnamefont
  {Soma}} \emph {et~al.},\ }\href {\doibase 10.1016/j.nima.2016.08.044}
  {\bibfield  {journal} {\bibinfo  {journal} {Nucl. Instrum. Methods Phys.
  Res., Sect. A}\ }\textbf {\bibinfo {volume} {836}},\ \bibinfo {pages} {67 }
  (\bibinfo {year} {2016})}\BibitemShut {NoStop}%
\bibitem [{\citenamefont {Cheng}\ \emph {et~al.}(2017)\citenamefont {Cheng}
  \emph {et~al.}}]{cjpl}%
  \BibitemOpen
  \bibfield  {author} {\bibinfo {author} {\bibfnamefont {J.~P.}\ \bibnamefont
  {Cheng}} \emph {et~al.},\ }\href {\doibase
  10.1146/annurev-nucl-102115-044842} {\bibfield  {journal} {\bibinfo
  {journal} {Annu. Rev. Nucl. Part. Sci.}\ }\textbf {\bibinfo {volume} {67}},\
  \bibinfo {pages} {231} (\bibinfo {year} {2017})}\BibitemShut {NoStop}%
\bibitem [{\citenamefont {She}\ \emph {et~al.}(2020)\citenamefont {She} \emph
  {et~al.}}]{cdex_dark+pthoton}%
  \BibitemOpen
  \bibfield  {author} {\bibinfo {author} {\bibfnamefont {Z.}~\bibnamefont
  {She}} \emph {et~al.} (\bibinfo {collaboration} {CDEX Collaboration}),\
  }\href {\doibase 10.1103/PhysRevLett.124.111301} {\bibfield  {journal}
  {\bibinfo  {journal} {Phys. Rev. Lett.}\ }\textbf {\bibinfo {volume} {124}},\
  \bibinfo {pages} {111301} (\bibinfo {year} {2020})}\BibitemShut {NoStop}%
\bibitem [{\citenamefont {Cesarsky}(1980)}]{cr_origin}%
  \BibitemOpen
  \bibfield  {author} {\bibinfo {author} {\bibfnamefont {C.~J.}\ \bibnamefont
  {Cesarsky}},\ }\href {\doibase 10.1146/annurev.aa.18.090180.001445}
  {\bibfield  {journal} {\bibinfo  {journal} {Annu. Rev. Astron. Astrophys.}\
  }\textbf {\bibinfo {volume} {18}},\ \bibinfo {pages} {289} (\bibinfo {year}
  {1980})}\BibitemShut {NoStop}%
\bibitem [{\citenamefont {Bykov}\ \emph {et~al.}(2018)\citenamefont {Bykov}
  \emph {et~al.}}]{cr_origin2}%
  \BibitemOpen
  \bibfield  {author} {\bibinfo {author} {\bibfnamefont {A.~M.}\ \bibnamefont
  {Bykov}} \emph {et~al.},\ }\href {\doibase 10.1007/s11214-018-0479-4}
  {\bibfield  {journal} {\bibinfo  {journal} {Space Sci. Rev.}\ }\textbf
  {\bibinfo {volume} {214}},\ \bibinfo {pages} {41} (\bibinfo {year}
  {2018})}\BibitemShut {NoStop}%
\bibitem [{\citenamefont {Aguilar}\ \emph {et~al.}(2014)\citenamefont {Aguilar}
  \emph {et~al.}}]{ams02}%
  \BibitemOpen
  \bibfield  {author} {\bibinfo {author} {\bibfnamefont {M.}~\bibnamefont
  {Aguilar}} \emph {et~al.} (\bibinfo {collaboration} {AMS Collaboration}),\
  }\href {\doibase 10.1103/PhysRevLett.113.121102} {\bibfield  {journal}
  {\bibinfo  {journal} {Phys. Rev. Lett.}\ }\textbf {\bibinfo {volume} {113}},\
  \bibinfo {pages} {121102} (\bibinfo {year} {2014})}\BibitemShut {NoStop}%
\bibitem [{\citenamefont {Yoon}\ \emph {et~al.}(2017)\citenamefont {Yoon} \emph
  {et~al.}}]{CREAM}%
  \BibitemOpen
  \bibfield  {author} {\bibinfo {author} {\bibfnamefont {Y.~S.}\ \bibnamefont
  {Yoon}} \emph {et~al.},\ }\href {\doibase 10.3847/1538-4357/aa68e4}
  {\bibfield  {journal} {\bibinfo  {journal} {Astrophys. J.}\ }\textbf
  {\bibinfo {volume} {839}},\ \bibinfo {pages} {5} (\bibinfo {year}
  {2017})}\BibitemShut {NoStop}%
\bibitem [{\citenamefont {An}\ \emph {et~al.}(2019)\citenamefont {An} \emph
  {et~al.}}]{DAMPE}%
  \BibitemOpen
  \bibfield  {author} {\bibinfo {author} {\bibfnamefont {Q.}~\bibnamefont {An}}
  \emph {et~al.},\ }\href {\doibase 10.1126/sciadv.aax3793} {\bibfield
  {journal} {\bibinfo  {journal} {Sci. Adv.}\ }\textbf {\bibinfo {volume}
  {5}},\ \bibinfo {pages} {eaax3793} (\bibinfo {year} {2019})}\BibitemShut
  {NoStop}%
\bibitem [{\citenamefont {Adriani}\ \emph {et~al.}(2019)\citenamefont {Adriani}
  \emph {et~al.}}]{CARLET}%
  \BibitemOpen
  \bibfield  {author} {\bibinfo {author} {\bibfnamefont {O.}~\bibnamefont
  {Adriani}} \emph {et~al.} (\bibinfo {collaboration} {CALET Collaboration}),\
  }\href {\doibase 10.1103/PhysRevLett.122.181102} {\bibfield  {journal}
  {\bibinfo  {journal} {Phys. Rev. Lett.}\ }\textbf {\bibinfo {volume} {122}},\
  \bibinfo {pages} {181102} (\bibinfo {year} {2019})}\BibitemShut {NoStop}%
\bibitem [{\citenamefont {Adriani}\ \emph {et~al.}(2011)\citenamefont {Adriani}
  \emph {et~al.}}]{pamela}%
  \BibitemOpen
  \bibfield  {author} {\bibinfo {author} {\bibfnamefont {O.}~\bibnamefont
  {Adriani}} \emph {et~al.},\ }\href {\doibase 10.1126/science.1199172}
  {\bibfield  {journal} {\bibinfo  {journal} {Science}\ }\textbf {\bibinfo
  {volume} {332}},\ \bibinfo {pages} {69} (\bibinfo {year} {2011})}\BibitemShut
  {NoStop}%
\bibitem [{\citenamefont {Cummings}\ \emph {et~al.}(2016)\citenamefont
  {Cummings} \emph {et~al.}}]{voyager}%
  \BibitemOpen
  \bibfield  {author} {\bibinfo {author} {\bibfnamefont {A.~C.}\ \bibnamefont
  {Cummings}} \emph {et~al.},\ }\href {\doibase 10.3847/0004-637x/831/1/18}
  {\bibfield  {journal} {\bibinfo  {journal} {Astrophys. J.}\ }\textbf
  {\bibinfo {volume} {831}},\ \bibinfo {pages} {18} (\bibinfo {year}
  {2016})}\BibitemShut {NoStop}%
\bibitem [{\citenamefont {Porter}\ \emph {et~al.}(2022)\citenamefont {Porter},
  \citenamefont {Johannesson},\ and\ \citenamefont {Moskalenko}}]{GALPROP_v57}%
  \BibitemOpen
  \bibfield  {author} {\bibinfo {author} {\bibfnamefont {T.~A.}\ \bibnamefont
  {Porter}}, \bibinfo {author} {\bibfnamefont {G.}~\bibnamefont {Johannesson}},
  \ and\ \bibinfo {author} {\bibfnamefont {I.~V.}\ \bibnamefont {Moskalenko}},\
  }\href {\doibase 10.3847/1538-4365/ac80f6} {\bibfield  {journal} {\bibinfo
  {journal} {Astrophys. J. Supp.}\ }\textbf {\bibinfo {volume} {262}},\
  \bibinfo {pages} {30} (\bibinfo {year} {2022})}\BibitemShut {NoStop}%
\bibitem [{\citenamefont {Boschini}\ \emph {et~al.}(2018)\citenamefont
  {Boschini} \emph {et~al.}}]{Boschini_2018}%
  \BibitemOpen
  \bibfield  {author} {\bibinfo {author} {\bibfnamefont {M.~J.}\ \bibnamefont
  {Boschini}} \emph {et~al.},\ }\href {\doibase 10.3847/1538-4357/aabc54}
  {\bibfield  {journal} {\bibinfo  {journal} {Astrophys. J.}\ }\textbf
  {\bibinfo {volume} {858}},\ \bibinfo {pages} {61} (\bibinfo {year}
  {2018})}\BibitemShut {NoStop}%
\bibitem [{\citenamefont {Padovani}\ \emph {et~al.}(2009)\citenamefont
  {Padovani}, \citenamefont {Galli},\ and\ \citenamefont
  {Glassgold}}]{molecular_cloud1}%
  \BibitemOpen
  \bibfield  {author} {\bibinfo {author} {\bibfnamefont {M.}~\bibnamefont
  {Padovani}}, \bibinfo {author} {\bibfnamefont {D.}~\bibnamefont {Galli}}, \
  and\ \bibinfo {author} {\bibfnamefont {A.~E.}\ \bibnamefont {Glassgold}},\
  }\href {\doibase 10.1051/0004-6361/200911794} {\bibfield  {journal} {\bibinfo
   {journal} {Astron. Astrophys.}\ }\textbf {\bibinfo {volume} {501}},\
  \bibinfo {pages} {619} (\bibinfo {year} {2009})}\BibitemShut {NoStop}%
\bibitem [{\citenamefont {{Padovani, Marco}}\ \emph {et~al.}(2022)\citenamefont
  {{Padovani, Marco}} \emph {et~al.}}]{molecular_cloud2}%
  \BibitemOpen
  \bibfield  {author} {\bibinfo {author} {\bibnamefont {{Padovani, Marco}}}
  \emph {et~al.},\ }\href {\doibase 10.1051/0004-6361/202142560} {\bibfield
  {journal} {\bibinfo  {journal} {Astron. Astrophys.}\ }\textbf {\bibinfo
  {volume} {658}},\ \bibinfo {pages} {A189} (\bibinfo {year}
  {2022})}\BibitemShut {NoStop}%
\bibitem [{\citenamefont {Navarro}\ \emph {et~al.}(1997)\citenamefont
  {Navarro}, \citenamefont {Frenk},\ and\ \citenamefont {White}}]{nfw}%
  \BibitemOpen
  \bibfield  {author} {\bibinfo {author} {\bibfnamefont {J.~F.}\ \bibnamefont
  {Navarro}}, \bibinfo {author} {\bibfnamefont {C.~S.}\ \bibnamefont {Frenk}},
  \ and\ \bibinfo {author} {\bibfnamefont {S.~D.~M.}\ \bibnamefont {White}},\
  }\href {\doibase 10.1086/304888} {\bibfield  {journal} {\bibinfo  {journal}
  {Astrophys. J.}\ }\textbf {\bibinfo {volume} {490}},\ \bibinfo {pages} {493}
  (\bibinfo {year} {1997})}\BibitemShut {NoStop}%
\bibitem [{\citenamefont {Bardhan}\ \emph {et~al.}(2023)\citenamefont
  {Bardhan}, \citenamefont {Bhowmick}, \citenamefont {Ghosh}, \citenamefont
  {Guha},\ and\ \citenamefont {Sachdeva}}]{energy_dependent}%
  \BibitemOpen
  \bibfield  {author} {\bibinfo {author} {\bibfnamefont {D.}~\bibnamefont
  {Bardhan}}, \bibinfo {author} {\bibfnamefont {S.}~\bibnamefont {Bhowmick}},
  \bibinfo {author} {\bibfnamefont {D.}~\bibnamefont {Ghosh}}, \bibinfo
  {author} {\bibfnamefont {A.}~\bibnamefont {Guha}}, \ and\ \bibinfo {author}
  {\bibfnamefont {D.}~\bibnamefont {Sachdeva}},\ }\href {\doibase
  10.1103/PhysRevD.107.015010} {\bibfield  {journal} {\bibinfo  {journal}
  {Phys. Rev. D}\ }\textbf {\bibinfo {volume} {107}},\ \bibinfo {pages}
  {015010} (\bibinfo {year} {2023})}\BibitemShut {NoStop}%
\bibitem [{\citenamefont {Emken}\ \emph {et~al.}(2019)\citenamefont {Emken}
  \emph {et~al.}}]{earthshielding4}%
  \BibitemOpen
  \bibfield  {author} {\bibinfo {author} {\bibfnamefont {T.}~\bibnamefont
  {Emken}} \emph {et~al.},\ }\href {\doibase 10.1088/1475-7516/2019/09/070}
  {\bibfield  {journal} {\bibinfo  {journal} {J. Cosmol. Astropart. Phys.}\
  }\textbf {\bibinfo {volume} {09}},\ \bibinfo {pages} {070} (\bibinfo {year}
  {2019})}\BibitemShut {NoStop}%
\bibitem [{\citenamefont {Bunge}\ \emph {et~al.}(1993)\citenamefont {Bunge},
  \citenamefont {Barrientos},\ and\ \citenamefont {Bunge}}]{noble_gas_table}%
  \BibitemOpen
  \bibfield  {author} {\bibinfo {author} {\bibfnamefont {C.}~\bibnamefont
  {Bunge}}, \bibinfo {author} {\bibfnamefont {J.}~\bibnamefont {Barrientos}}, \
  and\ \bibinfo {author} {\bibfnamefont {A.}~\bibnamefont {Bunge}},\ }\href
  {\doibase https://doi.org/10.1006/adnd.1993.1003} {\bibfield  {journal}
  {\bibinfo  {journal} {At. Data and Nucl. Data Tables}\ }\textbf {\bibinfo
  {volume} {53}},\ \bibinfo {pages} {113} (\bibinfo {year} {1993})}\BibitemShut
  {NoStop}%
\bibitem [{\citenamefont {Griffin}\ \emph {et~al.}(2021)\citenamefont
  {Griffin}, \citenamefont {Inzani}, \citenamefont {Trickle}, \citenamefont
  {Zhang},\ and\ \citenamefont {Zurek}}]{Rif1}%
  \BibitemOpen
  \bibfield  {author} {\bibinfo {author} {\bibfnamefont {S.~M.}\ \bibnamefont
  {Griffin}}, \bibinfo {author} {\bibfnamefont {K.}~\bibnamefont {Inzani}},
  \bibinfo {author} {\bibfnamefont {T.}~\bibnamefont {Trickle}}, \bibinfo
  {author} {\bibfnamefont {Z.}~\bibnamefont {Zhang}}, \ and\ \bibinfo {author}
  {\bibfnamefont {K.~M.}\ \bibnamefont {Zurek}},\ }\href {\doibase
  10.1103/PhysRevD.104.095015} {\bibfield  {journal} {\bibinfo  {journal}
  {Phys. Rev. D}\ }\textbf {\bibinfo {volume} {104}},\ \bibinfo {pages}
  {095015} (\bibinfo {year} {2021})}\BibitemShut {NoStop}%
\bibitem [{\citenamefont {Fernández-Serra}\ \emph {et~al.}(2016)\citenamefont
  {Fernández-Serra}, \citenamefont {Mardon}, \citenamefont {Soto},
  \citenamefont {Volansky},\ and\ \citenamefont {Yu}}]{Rif2}%
  \BibitemOpen
  \bibfield  {author} {\bibinfo {author} {\bibfnamefont {M.}~\bibnamefont
  {Fernández-Serra}}, \bibinfo {author} {\bibfnamefont {J.}~\bibnamefont
  {Mardon}}, \bibinfo {author} {\bibfnamefont {A.}~\bibnamefont {Soto}},
  \bibinfo {author} {\bibfnamefont {T.}~\bibnamefont {Volansky}}, \ and\
  \bibinfo {author} {\bibfnamefont {T.-T.}\ \bibnamefont {Yu}},\ }\href
  {\doibase 10.1007/JHEP05(2016)046} {\bibfield  {journal} {\bibinfo  {journal}
  {J. High Energy Phys.}\ }\textbf {\bibinfo {volume} {05}},\ \bibinfo {pages}
  {046} (\bibinfo {year} {2016})}\BibitemShut {NoStop}%
\bibitem [{\citenamefont {Li}\ \emph {et~al.}(2015)\citenamefont {Li},
  \citenamefont {Miao},\ and\ \citenamefont {Zhou}}]{Rif3}%
  \BibitemOpen
  \bibfield  {author} {\bibinfo {author} {\bibfnamefont {T.}~\bibnamefont
  {Li}}, \bibinfo {author} {\bibfnamefont {S.}~\bibnamefont {Miao}}, \ and\
  \bibinfo {author} {\bibfnamefont {Y.-F.}\ \bibnamefont {Zhou}},\ }\href
  {\doibase 10.1088/1475-7516/2015/03/032} {\bibfield  {journal} {\bibinfo
  {journal} {J. Cosmol. Astropart. Phys.}\ }\textbf {\bibinfo {volume} {03}},\
  \bibinfo {pages} {032} (\bibinfo {year} {2015})}\BibitemShut {NoStop}%
\bibitem [{\citenamefont {Trickle}\ and\ \citenamefont
  {kinzani}(2022)}]{exceed_dm}%
  \BibitemOpen
  \bibfield  {author} {\bibinfo {author} {\bibfnamefont {T.}~\bibnamefont
  {Trickle}}\ and\ \bibinfo {author} {\bibnamefont {kinzani}},\ }\href
  {\doibase 10.5281/zenodo.6097642} {\enquote {\bibinfo {title}
  {{tanner-trickle/EXCEED-DM: EXCEED-DM-v0.3.0}},}\ } (\bibinfo {year}
  {2022})\BibitemShut {NoStop}%
\bibitem [{\citenamefont {Bahcall}(1963)}]{K-X-ray}%
  \BibitemOpen
  \bibfield  {author} {\bibinfo {author} {\bibfnamefont {J.~N.}\ \bibnamefont
  {Bahcall}},\ }\href {\doibase 10.1103/PhysRev.132.362} {\bibfield  {journal}
  {\bibinfo  {journal} {Phys. Rev.}\ }\textbf {\bibinfo {volume} {132}},\
  \bibinfo {pages} {362} (\bibinfo {year} {1963})}\BibitemShut {NoStop}%
\bibitem [{\citenamefont {Barak}\ \emph {et~al.}(2020)\citenamefont {Barak}
  \emph {et~al.}}]{SENSEI}%
  \BibitemOpen
  \bibfield  {author} {\bibinfo {author} {\bibfnamefont {L.}~\bibnamefont
  {Barak}} \emph {et~al.} (\bibinfo {collaboration} {SENSEI Collaboration}),\
  }\href {\doibase 10.1103/PhysRevLett.125.171802} {\bibfield  {journal}
  {\bibinfo  {journal} {Phys. Rev. Lett.}\ }\textbf {\bibinfo {volume} {125}},\
  \bibinfo {pages} {171802} (\bibinfo {year} {2020})}\BibitemShut {NoStop}%
\bibitem [{\citenamefont {Aguilar-Arevalo}\ \emph {et~al.}(2019)\citenamefont
  {Aguilar-Arevalo} \emph {et~al.}}]{DAMIC}%
  \BibitemOpen
  \bibfield  {author} {\bibinfo {author} {\bibfnamefont {A.}~\bibnamefont
  {Aguilar-Arevalo}} \emph {et~al.} (\bibinfo {collaboration} {DAMIC
  Collaboration}),\ }\href {\doibase 10.1103/PhysRevLett.123.181802} {\bibfield
   {journal} {\bibinfo  {journal} {Phys. Rev. Lett.}\ }\textbf {\bibinfo
  {volume} {123}},\ \bibinfo {pages} {181802} (\bibinfo {year}
  {2019})}\BibitemShut {NoStop}%
\end{thebibliography}%

\end{document}